\documentclass[]{article}
\pdfoutput=1 

\usepackage{url}
\usepackage{libertine}
\usepackage{xcolor}
\definecolor{seasgreen}{HTML}{2e8b57}
\usepackage[colorlinks=true,
	linkcolor=violet,
	citecolor=seasgreen]{hyperref}
\usepackage{amssymb, amsfonts, amsmath, amsthm}
\usepackage{graphicx, natbib}
\usepackage[ruled,vlined]{algorithm2e}
\usepackage{accents}
\usepackage{enumitem}
\usepackage{authblk}

\newcommand{\BIT}{\begin{itemize}}
\newcommand{\EIT}{\end{itemize}}
\newcommand{\BNUM}{\begin{enumerate}}
\newcommand{\ENUM}{\end{enumerate}}

\newcommand\mbb[1]{\mathbb{#1}}

\def\reals{\mathbb{R}} 
\renewcommand{\exp}[1]{\operatorname{exp}\left(#1\right)} 
\def\indic#1{\mbb{I}\left({#1}\right)} 

\def\E{\mathbb{E}} 

\def\Eabsarg#1{\E\left|{#1}\right|}

\def\absarg#1{\left|#1\right|}

\theoremstyle{definition}
\newtheorem{definition}{Definition}[section]

\title{Measuring the Stability of Learned Features}
\author{Kris Sankaran}
\affil{Department of Statistics \\ University of Wisconsin - Madison \\ \href{mailto:ksankaran@wisc.edu}{ksankaran@wisc.edu}}

\begin{document}
\maketitle


\begin{abstract}
  Many modern datasets don't fit neatly into $n \times p$ matrices, but most
  techniques for measuring statistical stability expect rectangular data. We
  study methods for stability assessment on non-rectangular data, using
  statistical learning algorithms to extract rectangular latent features. We
  design controlled simulations to characterize the power and practicality of
  competing approaches. This motivates new strategies for visualizing feature
  stability. Our stability curves supplement the direct analysis, providing
  information about the reliability of inferences based on learned features.
  Finally, we illustrate our approach using a spatial proteomics dataset, where
  machine learning tools can augment the scientist's workflow, but where
  guarantees of statistical reproducibility are still central. Our raw data,
  packaged code, and experimental outputs are publicly available.
\end{abstract}


How should we perform statistical inference on nontabular data? One idea,
discussed in \citep{buhlmann2019comments}, is to first transform the data into
tabular form, taking advantage of modern feature learning algorithms. The point
is important enough to quote at length,

\begin{quote}
I would like to re-emphasize the importance of new sources of information.
Indeed, images, videos and audios are typically cheap devices to record data.
[The authors] do not mention recent progress with autoencoders
\citep{hinton2006reducing, vincent2010stacked}: when using such techniques, one
would again end up with numeric features which can then be used for further
downstream analysis using techniques from high-dimensional statistics or
statistical machine learning \citep{hastie2015statistical,
  buhlmann2011statistics}.
\end{quote}

Here, we explore this proposal in depth, illustrating how methods from
high-dimensional statistics can be used to study the inferential properties of
machine-generated features. Specifically, we present algorithms and
visualizations that can be used to characterize the statistical stability of
features learned from nontabular data sources. In the process, we also uncover
novel challenges particular to this new setting.

To ground the discussion, consider the following examples,

\begin{itemize}
\item Spatial Omics: In addition to measuring gene or protein expression at the
  cell level, it has become possible to study how expression varies spatially
  across a tissue \citep{lundberg2019spatial}. A variety of spatial features are
  thought to influence biological processes. For example, for some types of
  cancer, it is thought that the infiltration of tumors by immune cells
  expressing particular proteins can influence disease prognosis
  \citep{keren2018structured}.
\item Ecological and public health: Satellite imagery are now routinely
  integrated into ecological and public health projects, since they can capture
  important environmental features and are readily available
  globally \citep{rolf2020generalizable, bondimapping}. These methods have been
  found to be effective proxies for otherwise resource-intensive collection
  methods, like on-the-ground surveys, opening up the possibility of more
  universal and easily updated monitoring.
\end{itemize}

In both cases, machine learning methods are key for extracting useful features
from novel sources of data. However, unlike many common machine learning
applications, the learned features here are subject to critical examination,
either to inform biological mechanisms or to ensure vulnerable populations are
not put at risk.

This question of how to perform inference on learned features is not a new one —
principal components can be bootstrapped \cite{diaconis1983computer}, excess
error can be estimated from selected features \cite{gong1986cross}, and
confidence regions are available for exploratory projection pursuit
\cite{elguero1988confidence}. More recently, studies have investigated the use
of black-box predictions as features in downstream analysis
\citep{wang2020methods}.

However, for both the proteomics and satellite imagery examples presented above,
these methods can't be directly applied, because the raw data are not tabular.
Instead, features are typically automatically learned using deep learning. What
is new in this setting?

\begin{itemize}
\item Modern learned features are ``distributed'' \cite{mcclelland1986parallel}.
  This means that any pattern observed by the algorithm will be encoded by a
  large number of more elementary features, not any single feature specialized
  to that pattern. A deep learning algorithm is able to recognize a highway in a
  satellite image because it can merge evidence from across neurons (i.e., an
  elementary learned feature) that activate when particular color, shape, and
  edge features are present. This approach turns out to be much more effective
  than curating specialized features for many computational tasks, but it poses
  a challenge for human inspection.
\item From one run to the next, the learned features can change. This is unlike
  in principal components analysis, say, where the learned components are
  ordered in a natural way. If the deep learning features were more specialized,
  it might be possible to recognize the same feature across two runs, and then
  match them. However, the features are distributed, so it isn’t easy to say
  that any given neuron from the first run matches any other neuron(s) in the
  second.
\item It’s impractical to bootstrap methods that take hours to run, even if they
  could be done in parallel. Moreover, it’s unclear what information should be
  compared across bootstraps — the model parameters, the learned features, or
  summary statistics about the features.
\item Some form of sample-splitting must take place, to ensure that features are
  not evaluated using the same data that was used to learn them. However, it’s
  unclear how the splitting should be carried out. How much data should be used
  for feature learning, and how much for inference?
\end{itemize}

This work discusses these questions, proposing relevant definitions and
algorithms, examining their behavior through simulation, and illustrating their
use on a spatial proteomics dataset. Our basic takeaways are,

\begin{itemize}
\item While learned features are not interpretable when viewed in isolation,
  their associated feature subspaces often are. A feature learning algorithm may
  require a large number of elementary features in order to develop effective
  distributed representations, but the effective dimensionality of these
  representations is often small. These algorithms learn many, but highly
  correlated, features.
\item Given enough training data, learned feature subspaces are often stable
  from one run to the next, and this can be quantified using a Procrustes
  analysis. Unsupervised feature learning algorithms are typically more stable
  than supervised ones.
\item For problems where unsupervised feature learning is effective, a fast
  approximation to full deep model training, called the random convolutional
  features (RCF) model, can suffice for a feature stability analysis.
\item Inference is less data-greedy than feature learning, in the sense that
  when few samples are reserved for inference, stable features can still be
  identified. This is no longer the case when few samples are reserved for
  feature learning.
\end{itemize}

Section \ref{sec:psetup} provides a description of our problem setting. Section
\ref{sec:context} summarizes the key technical tools used in this study. We
present generic algorithms for measuring feature subspace stability in Section
\ref{sec:algorithms}, and we study its properties through a simulation in
Section \ref{sec:simulation}. We conclude in Section \ref{sec:dataset} with an
application to a spatial proteomics dataset.

\section{Problem Setup}
\label{sec:psetup}

Our goal is to characterize the stability of features that were algorithmically
derived from $n$ samples $x_i \in \mathcal{X}$. For example, each $x_{i}$ might
be a spatial proteomics measurement or a satellite image. They could also be
more general data types -- $x_i$ might be the audio recording for one of $n$
speakers, or the graph derived from one of $n$ molecules. Optionally, a vector
$\mathbf{y}\in \reals^{n}$ of responses associated with each observation will be
available. We denote the full set of available data by $\mathcal{D}$.

\begin{definition}
A \textit{feature learner} is a parameterized mapping $T\left(\cdot;
\theta\right): \mathcal{X} \to \reals^{L}$ which takes data from the raw input
domain $\mathcal{X}$ and represents it using a vector in $\reals^{L}$.
\end{definition}

For example, in the proteomics and satellite applications, we expect the learner
to transform a set of raw pixel intensities into a vector of features reflecting
biological or environmental properties of the imaged sample. The parameter
$\theta$ is estimated from data, typically through an optimization problem,
\begin{align*}
  \hat{\theta} := \arg\min_{\theta \in \Theta} \mathcal{L}\left(\mathcal{D}, T\left(\cdot; \theta\right)\right)
\end{align*}
for some loss function $\mathcal{L}$. In an unsupervised feature learner,
candidates $\theta \in \Theta$ are functions of $x_{1}, \dots, x_{n}$ alone. For
a supervised feature learner, the class includes functions of both
$x_{1}, \dots, x_{n}$ and $\mathbf{y}$. To simplify notation, we will write
$z_{i} = T\left(x_{i}; \hat{\theta}\right) \in
\reals^{L}$ to denote the learned features associated with the $i^{th}$
observation.

A first attempt might try to assign a stability score to each of the $L$
coordinates of $z$. An investigator would then have confidence that, if data
were sampled again from the same population, and if features were extracted by
the same black box, then the features with high stability scores would reappear
in the new analysis. The essential challenge is that the learned features are
not the same from one run to the next; the $l^{th}$ learned feature from run 1
need not have any relationship with the $l^{th}$ feature from run 2. Parallel
problems are well-known in clustering and factor analysis. Cluster labels can be
permuted without changing the quality of the clustering, leading to the
label-switching problem. Likewise, without any criteria for sorting or
postprocessing factors, latent factors from a factor analysis are
unidentifiable.

To address this issue, we distinguish between two notions of stability, which we
call feature subspace and selection stability, respectively. The idea of
subspace stability is that, even if there is no direct correspondence between
learned features across runs, they may all reflect the same underlying latent
features. Different runs of the feature learning algorithm return different
bases for nearly the same subspace. To make this more precise, suppose that the
$b^{th}$ run of the feature learning algorithm produces,

\begin{align*}
\mathbf{Z}_{b} &= \begin{pmatrix}
z^{b}_{1} \\
\vdots \\
z^{b}_{n}
\end{pmatrix} \in \reals^{n \times L}
\end{align*}

and define an alignment function $\mathcal{A}$ which takes learned to aligned
features,
\begin{align*}
\mathcal{A}: \mathbf{Z}_{1}, \dots, \mathbf{Z}_{B} \to \mathbf{M}, \left(\underaccent{\bar}{\mathbf{Z}}_{1}, \dots, \underaccent{\bar}{\mathbf{Z}}_{B}\right).
\end{align*}

We think of $\mathbf{M} \in \reals^{n \times K}$ as the average of all $B$
representations and $\underaccent{\bar}{\mathbf{Z}}_{b} \in \reals^{n \times K}$
as the version of the $b^{th}$ learned representation after they have been put
into a common coordinate system. $\mathcal{A}$ is just a Procrustes analysis.
With this notation, we can now define subspace stability.
\begin{definition}
The \textit{subspace stability} of $B$ learned representations $\mathbf{Z}_{1},
\dots, \mathbf{Z}_{B}$ with respect to an alignment function $\mathcal{A}$ is
the average distance between the aligned features and $\mathbf{M}$,
\begin{align*}
FSS_{\mathcal{A}}\left(\mathbf{Z}_{1}, \dots, \mathbf{Z}_{B}\right) &= \frac{1}{B} \sum_{b = 1}^{B} \|\underaccent{\bar}{\mathbf{Z}}_{b} - \mathbf{M}\|^{2}_{2} \\
\text{where } \mathbf{M}, \left(\underaccent{\bar}{\mathbf{Z}}_{1}, \dots, \underaccent{\bar}{\mathbf{Z}}_{B}\right) &= \mathcal{A}\left(\mathbf{Z}_{1}, \dots, \mathbf{Z}_{B}\right)
\end{align*}
\end{definition}

By selection stability, we mean that a given aligned feature is repeatedly
discovered by a model selection procedure. At this point, the features can be
thought of as fixed; we are back on more familiar inferential ground. That is,
let $\mathcal{S}$ be a selection function, which takes
$\underaccent{\bar}{\mathbf{Z}}_{b}$ and a response $\mathbf{y}$ and returns a
subset $S_{b} \subseteq \{1, \dots, K\}$ of features important for predicting
$\mathbf{y}$.

\begin{definition}
The \textit{selection stability} of the $k^{th}$ aligned feature with respect to
the selection $\mathcal{S}$ is the
fraction
\begin{align*}
SS_{\mathcal{S}}^{k}\left(\underaccent{\bar}{\mathbf{Z}}_{1}, \dots, \underaccent{\bar}{\mathbf{Z}}_{B}\right) &= \frac{1}{B}\sum_{b = 1}^{B} \indic{k \in \mathcal{S}\left(\underaccent{\bar}{\mathbf{Z}}_{b}, \mathbf{y}\right)}.
\end{align*}
\end{definition}


\section{Background}
\label{sec:context}

We review techniques that are used in our algorithmic proposals and
computational experiments.

\subsection{Stability}

Many statistical quantities are based on the idea that meaningful conclusions
should be stable to perturbations \cite{yu2013stability}. We will make use of
stability selection, a technique for evaluating significance in high-dimensional
linear models \cite{meinshausen2010stability}.

For a dataset $\mathbf{X}\in \reals^{n \times p}$ and $\mathbf{y} \in
\reals^{n}$, stability selection proceeds as follows. First, $B$ subsamples of
size $\lfloor \frac{n}{2} \rfloor$ are drawn from the data, and for each a lasso
regression is run over a grid of regularization parameters $\lambda$. Each of
these regressions results in $p$ coefficient trajectories
$\hat{\beta}_{j}\left(\lambda\right)$, and important features are expected to
become nonzero earlier on the regularization path (that is, even with large
regularization $\lambda$). For a given $\lambda$ and feature $j$, let
$\hat{\Pi}_{j}\left(\lambda\right)$ measure the fraction of subsamples for which
$\absarg{\hat{\beta}_j\left(\lambda\right)} > 0$. The paths
$\left(\hat{\Pi}_{j}\left(\lambda\right)\right)_{j = 1}^{p}$ describe the
importance of each of the $p$ regression features. For a given stringency
threshold $\pi_{thr}$, the procedure selects features $\hat{S} = \{j :
\max_{\lambda} \hat{\Pi}_{j}\left(\lambda\right) \geq \pi_{thr}\}$.

Let $q_{\Lambda} = \Eabsarg{\hat{S}_{\Lambda}}$ denote the expected number of
selected features. It can be shown that, for $\pi_{thr} \geq \frac{1}{2}$, and
assuming a sufficiently well-behaved $\hat{S}$, the expected number of falsely
selected features is bounded by $\frac{1}{2\pi_{thr} - 1}
\frac{q_{\Lambda}^2}{p}$. The term $\frac{q_{\Lambda}}{p}$ is like the typical
fraction of selected features; the coefficient $\frac{q_{\Lambda}}{2\pi_{thr} -
  1}$ describes the fraction of those that are expected to be false positives.

\subsection{Feature Learning}

We consider three particular feature learning algorithms. The
first is called the Variational Autoencoder (VAE). Like principal components
analysis, this algorithm learns an $L$-dimensional representation of a dataset
by optimizing a reconstruction objective. Formally, the algorithm first posits a
generative model $p\left(z\right)p_{\xi}\left(x \vert z\right)$ of the data;
$p\left(z\right)$ is a prior on latent features and $p_{\xi}\left(x \vert
z\right)$ is a likelihood parameterized by $\xi$. The algorithm finds a pair
$\xi, \varphi$ maximizing the lower bound,
\begin{align*}
\log p_{\xi}\left(x\right) \geq  \mathbb{E}_{q_{\varphi}}\left[\log p_{\xi}(x \mid z)\right]-D_{KL}\left(q_{\varphi}(z \mid x) \| p(z)\right)
\end{align*}
where $q_{\varphi}\left(z \vert x\right) = N\left(\mu_{\varphi}\left(x\right),
\sigma^{2}_{\varphi}\left(x\right)\right)$ maps raw data examples to
distributions in a latent space of more meaningful features. This optimization
problem is nonconvex, and is typically solved through a variant of stochastic
gradient descent. There are many particular implementations of the VAE; our
experiments are based on \citep{van2017neural}.

Second, we investigate features learned in a supervised way through a
Convolutional Neural Network (CNN). For regression, a CNN optimizes the
empirical estimate of the risk $\mathbf{E}\|y -
f_{W_{1:J}}\left(x\right)^{T}\beta\|_{2}^{2}$ over $W_{1:J}$ and $\beta$.
$f_{W_{1:J}}$ transforms the raw input into the ``last layer''’s features, and
is defined recursively according to
\begin{align*}
f^{j}_{W_{1:j}}\left(x\right) &= \sigma\left(W_{j}f^{j - 1}_{W_{1:(j - 1)}}\left(x\right)\right)\\
f^{0}\left(x\right) &= x
\end{align*}
for $\sigma\left(x\right)$ defined as $\sigma\left(x\right) := x \indic{x \geq
  0}$ and matrices $W$ restricted to the set of convolution operators. Like in
the VAE, this optimization is nonconvex and is typically found through
first-order optimization methods. Our particular implementation is the CBR
architecture described in \citep{raghu2017svcca}.

Third, we consider a random convolutional features (RCF) model
\citep{rahimi2008weighted}. A random sample of $L$ training examples
$x_{i_1}, \dots, x_{i_L} \in \reals^{w \times h \times c}$ are selected; the
$x_{i}$'s are assumed to be $c$-channel images with dimension $w\times h$. For
each sample, a random $s \times s$ patch, which we call $w_{p} \in \reals^{s
  \times s \times c}$, is extracted. For any $c$-channel image $x$, the $l^{th}$
feature $z_{l}$ is found by convolving $x$ with $w_{l}$ and spatially averaging
over all activations.

To train an RCF, the training data are first featurized into $\mathbf{Z} \in
\reals^{n \times L}$ using random image patches, as described above. Then, a
ridge regression model is trained, solving
\begin{align*}
\hat{\beta} := \arg \min_{\beta} \|y - \mathbf{Z}\beta\|_{2}^{2} + \lambda \|\beta\|_{2}
\end{align*}
For a new example $x^{\ast}$, the same image patches $w_{1}, \dots, w_{L}$ are
used to create a featurization $z^{\ast}$, and predictions are made with
$z^{\ast T}\hat{\beta}$. Unlike either the VAE or CNN, this model does not
require gradient based training, and it can serve as a fast, and often
effective, baseline.

\subsection{Procrustes Analysis}
Procrustes Analysis gives a way of aligning multiple tables. Given two centered
tables $\mathbf{X}$ and $\mathbf{Y}$, the Procrustes problem finds the rotation
matrix $\mathbf{R}$ solving the optimization,
\begin{align*}
\min_{\mathbf{R} \in \mathcal{O}\left(p, p\right)} \|\mathbf{X} - \mathbf{Y}\mathbf{R}\|^{2}_{F},
\end{align*}
where $\mathcal{O}\left(p, p\right)$ denotes the space of $p\times p$
orthonormal matrices. Using the associated Lagrangian,
\begin{align*}
\max_{\mathbf{R}, \mathbf{\Lambda} \succeq 0} \|\mathbf{X} - \mathbf{Y}\mathbf{R}\|_{F}^{2} - \mathbf{\Lambda}\left(\mathbf{R}^{T}\mathbf{R} - \mathbf{I}\right)
\end{align*}
the solution can be shown to be $\hat{\mathbf{R}} = \mathbf{U}^{T}\mathbf{V}$
for $\mathbf{U}$ and $\mathbf{V}$ obtained by the SVD of
$\mathbf{X}^{T}\mathbf{Y}$. For $B$ matrices $\mathbf{X}_{1}, \dots,
\mathbf{X}_{B}$, the generalized Procrustes problem finds $B$ rotations
$\mathbf{R}_{1}, \dots, \mathbf{R}_{B}$ and mean matrix $\mathbf{M}$ solving
\begin{align*}
\min_{\mathbf{R}_{1}, \dots, \mathbf{R}_{B} \in \mathcal{O}\left(p, p\right), M} \sum_{b = 1}^{B} \|\mathbf{X}_{b}\mathbf{R}_{b} - \mathbf{M}\|_{F}^{2}.
\end{align*}
While there is no closed form solution, the optimization can be solved by
cyclically updating each $\mathbf{R}_{b}$ via standard Procrustes problems and
then updating $\mathbf{M} = \frac{1}{B} \sum_{b = 1}^{B} \mathbf{X}_{b}
\mathbf{R}_{b}$.

\subsection{Representation Analysis}

Our approach is closely related to Singular Vector Canonical Correlation
Analysis (SVCCA), a method used to compare features learned across deep learning
models. In SVCCA, the two representations are identified with two matrices, $X
\in \reals^{n \times l_{1}}$ and $Y \in \reals^{n \times l_{2}}$. The $ij^{th}$
cell in each matrix corresponds to the activation of neuron $j$ on sample $i$. A
representation is a pattern of activations across a collection of neurons.

To compare representations, SVCCA first computes singular value decompositions
$X = U_{X}D_{X}V_{X}^{T}$ and $Y = U_{Y}D_{Y}V_{Y}^{T}$. The coordinates of
these matrices with respect to the top $K$ singular vector directions are then
extracted, $Z_{X} = U_{X, 1:K}D_{X, 1:K}$ and $Z_{Y} = U_{Y, 1:K}D_{Y, 1:K}$.
Finally, a canonical correlation analysis is performed on these coordinate
matrices. That is, the top CCA directions $u_{1}, v_{1}$ are found by optimizing
\begin{align*}
  \text{maximize }& u^{T}\hat{\Sigma}_{Z_{X}}^{-\frac{1}{2}}\hat{\Sigma}_{Z_{X}Z_{Y}}\hat{\Sigma}_{Z_{Y}}^{-\frac{1}{2}}v \\
  \text{subject to } & \|u\|_{2} = \|v\| = 1
\end{align*}
and subsequent directions are found by solving the same problem after
orthogonalizing $Z_{X}$ and $Z_{Y}$ to previously found directions. The value of
the objective for each of the $K$ directions is denoted $\rho_{k}$, and the
overall similarity between the two representations is taken to be the average of
these canonical correlations: $\frac{1}{K}\sum_{k = 1}^{K} \rho_{k}$.

Note that, while in principle, it would be possible to perform a CCA on the
activations $X$ and $Y$ directly, for representations with many neurons, the
dimensionality reduction step can simplify computation, because it avoids
inverting a high-dimensional covariance matrix.

\subsection{Sparse Components Analysis}

As in SVCCA, we reduce the dimensionality of learned features before comparing
them. We use both PCA and a variant called Sparse Components Analysis (SCA)
\citep{chen2020new}. For a given data matrix $X$ and a number of components $K$,
and sparsity parameter $\gamma$, SCA solves the optimization
\begin{align*}
  \text{maximize }_{Z, B, Y} &\|X - Z B Y^{T}\|_{F} \\
  \text{subject to }Z \in &\mathcal{O}\left(n, K\right) \\
  Y \in &\mathcal{O}\left(p, k\right) \\
  \|Y\|_{1} &\leq \gamma.
\end{align*}
The matrix $Y$ provides the $K$ SCA loadings, and the product $ZB$ provide the
coordinates of each sample with respect to that basis. Note that if $B$ is
forced to be diagonal, then this optimization reduces to sparse PCA. This
optimization does not directly provide an ordering of the loadings. However, the
proportion of variance explained by each dimension can still be computed, and
this can be used to re-order the dimensions in a similar way to PCA.


\section{Algorithms}
\label{sec:algorithms}

Algorithms \ref{alg:features} through \ref{alg:selection} give our approach to
measuring feature stability. We first motivate the general setup, before
explaining specific choices used in our experiments.

The three algorithms do the following,
\begin{enumerate}
  \item Train a feature learning algorithm on $B$ perturbed versions of the
    dataset $\mathcal{D}$. This yields $B$ sets of learned features
    $\mathbf{Z}_{b}$.
  \item Use an alignment strategy $\mathcal{A}$ to represent the learned
    features in a shared coordinate system. These aligned features are called
    $\underaccent{\bar}{\mathbf{Z}}_{b}$.
  \item Using the aligned features, evaluate the importance of each feature
    dimension using a selection mechanism $\mathcal{S}$.
\end{enumerate}

The only subtlety is that the feature learning and selection steps are performed
on different subsets of $\mathcal{D}$, indexed by $I$ and $I^{C}$, respectively.
This is needed to maintain validity of inference -- if the same samples were
used for selection and learning, features would appear more important than they
are.

In our experiments, we use the bootstrap to perturb the data reserved for
feature learning. That is, if $I_{b}^{\ast}$ resamples $I$ with replacement,
then $\mathcal{D}\left[I^{\ast}_{b}\right]$ gives a draw from
$\mathcal{P}\left(\mathcal{D}\right)$. This lets us obtain $B$ sets of learned
features by optimizing
\begin{align*}
  \hat{\theta}_{b} := \arg\min_{\theta \in \Theta} \mathcal{L}\left(\mathcal{D}\left[I_{b}^{\ast}\right], T\left(\cdot; \theta\right)\right)
\end{align*}
and setting $\mathbf{Z}_{b} = T\left(\mathcal{D}, \hat{\theta}_{b}\right)$. This
step is summarized by Algorithm \ref{alg:features}.

\begin{algorithm}[H]
\SetAlgoLined
\KwResult{Sets of learned features $\left(\mathbf{Z}_{b}\right)_{b = 1}^{B}$ for
  $\mathcal{D}$. Indices $I\subset\left[n\right]$ used for feature learning.}
Inputs: Dataset $\mathcal{D}$ and perturbation process $\mathcal{P}$. Candidate
feature learners $\{T\left(\cdot; \theta\right)\}_{\theta \in \Theta}$ and
criterion $\mathcal{L}$.

1. Randomly split samples into disjoint subsets $I, I^{C} \subset \left[n\right]$.

2. Generate $B$ perturbed datasets,

\For{$b = 1, \dots, B$}{
	$\mathcal{D}_{b} \sim \mathcal{P}\left(\mathcal{D}\left[I\right]\right)$
}

3. Train $B$ feature learners,

\For{$b = 1, \dots, B$}{
  $\hat{\theta}_{b} = \arg\min_{\theta \in \Theta}\mathcal{L}\left(\mathcal{D}_{b}, T\left(\cdot, \theta\right)\right)$ \\
  $\mathbf{Z}_{b} = T\left(\mathcal{D}, \hat{\theta}_{b}\right)$
}
\caption{Feature Learning}
\label{alg:features}
\end{algorithm}

For the alignment strategy $\mathcal{A}$, we first reduce the dimensionality of
each $\mathbf{Z}_{b}$ to $K$ dimensions, call this $\tilde{\mathbf{Z}}_{b}$.
Then, we solve a generalized Procrustes problem, finding $\mathbf{R}_{b}$'s so
that the $\underaccent{\bar}{\mathbf{Z}}_{b} :=
\tilde{\mathbf{Z}}_{b}\mathbf{R}_{b}$ have low Frobenius distance to
$\mathbf{M}$. For the dimensionality reduction step, we apply either PCA or SCA
after centering and scaling. Given this $\mathcal{A}$, we can compute a feature
subspace stability score using Algorithm \ref{alg:fss}.

\begin{algorithm}[H]
  \SetAlgoLined
  \KwResult{Aligned features $\left(\underaccent{\bar}{\mathbf{Z}}_{1}, \dots,
    \underaccent{\bar}{\mathbf{Z}}_{B}\right)$. Subspace stability score $FSS_{\mathcal{A}}$.}
  Inputs: Learned features $\left(\mathbf{Z}_{1}, \dots, \mathbf{Z}_{B}\right)$.
  Alignment strategy $\mathcal{A}$.

  1. Align features,
  \begin{align*}
    M, \left(\underaccent{\bar}{\mathbf{Z}}_{1}, \dots, \underaccent{\bar}{\mathbf{Z}}_{B}\right) = \mathcal{A}\left(\mathbf{Z}_{1}, \dots, \mathbf{Z}_{B}\right)
  \end{align*}

  2. Compute feature subspace stability,
  \begin{align*}
    FSS_{\mathcal{A}}\left(\mathbf{Z}_{1}, \dots, \mathbf{Z}_{B}\right) = \frac{1}{B} \sum_{b = 1}^{B} \|\underaccent{\bar}{\mathbf{Z}}_{b} - \mathbf{M}\|^{2}_{2}.
  \end{align*}
  \caption{Feature Subspace Stability}
  \label{alg:fss}
\end{algorithm}

For the selection mechanism, we use stability selection. This means that our
selection mechanism $\mathcal{S}$ is parameterized by lasso regularization
strength $\lambda$ and selection stringency $\pi_{thr}$. In our experiments, we
display the full selection curves $\hat{\Pi}^{b}_{k}\left(\lambda\right)$ for
each set of aligned features. From these curves, we can identify important
features $S_{b}\left(\lambda, \pi_{thr}\right)$ for any choice of $\lambda$ or
$\pi_{thr}$. However, for clarity, we suppress this dependence on $\lambda$ and
$\pi_{thr}$ in Algorithm \ref{alg:selection}.

\begin{algorithm}[H]
  \SetAlgoLined
  \KwResult{Selection stability scores for all aligned dimensions $SS_{\mathcal{S}}^{1}, \dots, SS_{\mathcal{S}}^{K}$.}
  Inputs: Reserved indices $I^{C}$. Selection mechanism $\mathcal{S}$. Aligned
  features $\left(\underaccent{\bar}{\mathbf{Z}}_{1}, \dots,
  \underaccent{\bar}{\mathbf{Z}}_{B}\right)$, where each
  $\underaccent{\bar}{\mathbf{Z}}_{b} \in \reals^{n \times K}$.

  1. Define selection sets,

  \For{$b = 1, \dots, B$}{
    $S_{b} = \mathcal{S}\left(\underaccent{\bar}{\mathbf{Z}}_{b}\left[I^{C}\right]\right)$
  }

  2. Compute selection scores,

  \For{$k = 1, \dots, K$}{
    $SS_{\mathcal{S}}^{k} = \frac{1}{B}\sum_{b = 1}^{B} \indic{k \in S_{b}}$
  }
  \caption{Selection Stability}
  \label{alg:selection}
\end{algorithm}


\section{Simulation Experiment}
\label{sec:simulation}

We use a simulation experiment to understand properties of the proposed
algorithms. Our guiding questions are,

\begin{enumerate}[label=(G{{\arabic*}})]
\item How different are the results obtained via supervised vs. unsupervised
  feature learning algorithms $T\left(\cdot; \theta\right)$?
\item When we vary the relative sizes of $I$ and $I^{C}$, we expect a trade-off
  between feature learning and inferential quality. Can we characterize this
  trade-off?
\item How does the dimensionality reduction approach used in $\mathcal{A}$
  affect downstream inferences?
\item Is it ever possible to substitute the RCF feature learner for either of
  the more time-consuming VAE or CNN models?
\end{enumerate}

To answer these questions, we evaluate our proposal using a full factorial
design with three factors,

\begin{enumerate}
\item Relative sizes of $I$ and $I^{C}$: We use $\absarg{I} \in \left\{0.15n,
  0.5n, 0.9n\right\}$.
\item Dimensionality reduction procedure: We use both PCA and SCA.
\item Algorithm used: We train CNN, VAE and RCF models.
\end{enumerate}

We use $B = 20$ perturbations in each case. Hence, 3 splits $\times$ 3 models
$\times$ 20 perturbations = 180 models are trained, from which 180 $\times$ 2
reductions $= 360$ dimensionality-reduced features are obtained.

\subsection{Simulated Data}

The key ingredient in this simulation is the construction of a dataset where all
``true'' features are directly controlled. To motivate the simulation, imagine
studying a set of tumor pathology slides, with the hope of discovering features
that are predictive of disease prognosis. Each pathology slide gives a view into
a cellular ecosystem — there are several types of cells, with different sizes,
density, and affinities to one another.

\begin{figure}
  \centering
  \includegraphics[width=\textwidth]{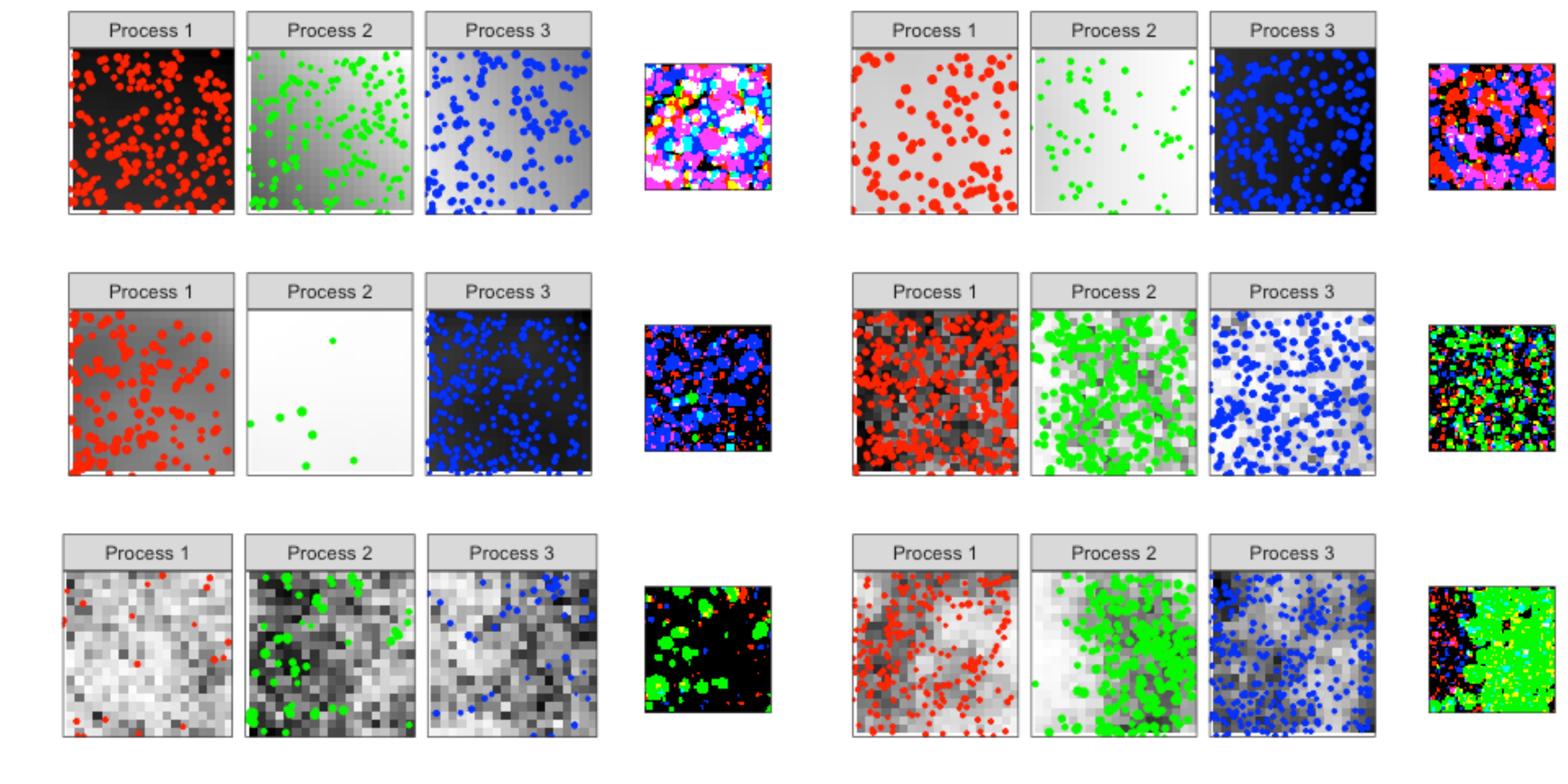}
  \caption{Example simulated images, for low (top), average (middle), and high
    (bottom) values of $y_i$. For each sample, three relative intensity
    functions $B_{k}\left(x\right)$ are generated, shown as a greyscale
    heatmaps. Samples drawn from each process are overlaid as circles. The final
    images, displayed at the right of each group, combines points from the three
    processes, removing the original generating intensity function, which is not
    available to the feature learner. Small $y_i$ values are associated with
    smoother, less structured intensity functions.}
  \label{fig:matern_example}
\end{figure}

In the simulation, we first generate the latent properties of each slide.
Prognosis is defined as a linear function of these properties. Then, images are
generated that also reflect the latent properties. The essential challenge is
that the investigator only has access to the images and patient prognoses, not
the true properties behind each image. A good feature extractor $T\left(x;
\hat{\theta}\right)$ should recover important cell ecosystem properties from the
images alone. Example images for varying values of $y$ are given in Figure
\ref{fig:matern_example}. In total, our simulation generates 10,000 such RGB
images, each of dimension $64 \times 64 \times 3$.

We now give details. The locations of cells are governed by an intensity
function drawn from a two-dimensional marked Log Cox Matern Process (LCMP)
\cite{diggle2013spatial}. Recall that a Matern process is a Gaussian process
with covariance function,
\begin{align}
  \label{eq:cov_lcmp}
C_{\nu, \alpha}(\|x - y\|)=\sigma^{2} \frac{2^{1-\nu}}{\Gamma(\nu)}\left(\sqrt{2 \nu} \frac{\|x - y\|}{\alpha}\right)^{\nu} K_{\nu}\left(\sqrt{2 \nu} \frac{\|x - y\|}{\alpha}\right),
\end{align}
where $\alpha$ acts like a bandwidth parameter and $\nu$ controls the roughness
of the simulated process.

Suppose we have $R$ types of cells. Then, our LCMP should have $R$ classes. This
can be constructed as follows. First, a nonnegative process
$\Lambda\left(x\right)$ is simulated along the image grid,
$\Lambda\left(x\right) \sim \exp{\mathcal{N}\left(0, \mathbf{C}_{\nu_{\Lambda},
    \alpha_{\Lambda}}\right)}$, where $\mathbf{C}_{\nu_{\Lambda},
  \alpha_{\Lambda}}$ is the covariance matrix induced by the $C_{\nu_{\Lambda},
  \alpha_{\Lambda}}$ in equation \ref{eq:cov_lcmp}. This is a baseline intensity
that determines the location of cells, regardless of cell type. Then, $R$
further processes are simulated, $B_{r}\left(x\right) \sim \exp{\beta_{r} +
  \mathcal{N}\left(0, \mathbf{C}_{\nu_{B}, \alpha_{B}}\right)} $. These
processes will reflect the relative frequencies of the $R$ classes at any
location $x$; the intercept $\beta_r$ makes a class either more or less frequent
across all positions $x$.

Given these intensity functions, we can simulate $N$ cell locations by drawing
from an inhomogeneous Poisson process with intensity $\Lambda\left(x\right)$.
For a cell at location $x$, we assign it cell type $r$ with probability
$\frac{B_{r}^{\tau}\left(x\right)}{\sum_{r^\prime = 1}^{R}
  B^{\tau}_{r^\prime}\left(x\right)}$. Here, we have introduced a temperature
parameter $\tau$ which controls the degree of mixedness between cell types at a
given location.

To complete the procedure for simulating images, we add two last source of
variation — the number of cells and the cell size. The number of cells per image
is drawn uniformly from 50 to 1000. The cells from class $R$ are drawn with a
random radius drawn from $\text{Gamma}\left(5, \lambda_{r}\right)$. A summary of
all parameters used to generate each image is given in Table
\ref{tab:sim_params}. Each parameter is drawn uniformly within its range, which
has been chosen to provide sufficient variation in image appearance. These
parameters are the ``true'' latent features associated with the simulated
images; they give the most concise description of the variation observed across
the images.

\begin{table}[]
\begin{tabular}{|p{0.1\linewidth}|p{0.4\linewidth}|p{0.16\linewidth}|p{0.15\linewidth}|}
\hline
\textbf{Feature}              & \textbf{Description}                                 & \textbf{Influence}          & \textbf{Range}             \\
\hline
$N_i$                & The total number of cells.                                    & 0.5                         & $\left[50, 1000\right]$    \\
\hline
$\nu_{i,\Lambda}$    & The roughness of the overall intensity process.            & -0.5                        & $\left[0, 8\right]$        \\
\hline
$\alpha_{i,\Lambda}$ & The bandwidth of the overall intensity process.            & -0.5                        & $\left[0, 8\right]$        \\
\hline
$\beta_{ir}$             & The intercept controlling the frequency of class $r$. & 1 for $r = 1$,\ -1 otherwise & $\left[-0.15, 0.15\right]$ \\
\hline
$\nu_{iB}$           & The roughness of the relative intensity processes.            & -0.5                        & $\left[0, 3\right]$        \\
\hline
$\alpha_{iB}$        & The bandwidth of relative intensity processes.                & -0.5                        & $\left[0, 3\right]$        \\
\hline
$\tau_{i}$           & The temperature used in cell type assignment.                 & 0.5                         & $\left[0, 3\right]$        \\
\hline
$\lambda_{ir}$       & The shape parameter controlling the sizes of each cell type.  & 1 for $r = 1$,\ 0 otherwise  & $\left[100, 500\right]$   \\
\hline
\end{tabular}
\caption{Our simulation mechanism is governed by the above parameters.
  Parameters $N_i$ and $\lambda_{ir}$ control the number and sizes of imaginary
  cells. $\nu_{i, \Lambda}$, $\alpha_{i, \Lambda}$, $\beta_{ik}$, $\nu_{iB}$,
  $\alpha_{iB}$, and $\tau_{i}$ control the overall and relative intensities of
  the marked LCMP from which the cell positions are drawn. Example draws are
  displayed in Figure \ref{fig:matern_example}.}
\label{tab:sim_params}
\end{table}

These features are the latent properties used to generate the prognosis for
each patient $i$. Specifically, we generate $y_i = \sum_{k} \text{Influence}_{k}
\times \left[\frac{\text{Feature}_{ik} -
    \overline{\text{Feature}}_{k}}{SD\left(\text{Feature}_{k}\right)}\right]$ for the
values Influence given in Table \ref{tab:sim_params}. Note that there is no
additional noise: if the $\text{Feature}_{ik}$'s were known for each sample,
then the $y_{i}$'s could be predicted perfectly. Therefore, the simulation
gauges the capacity of the feature learners to recover these known latent
features.

\subsection{Results}

We now summarize findings of our factorial experiment. All simulated data,
trained models, and aligned features have been posted publicly; links can be
found in the appendix.

\subsubsection{Distributed features}

\begin{figure}
  \centering
  \includegraphics[width=\textwidth]{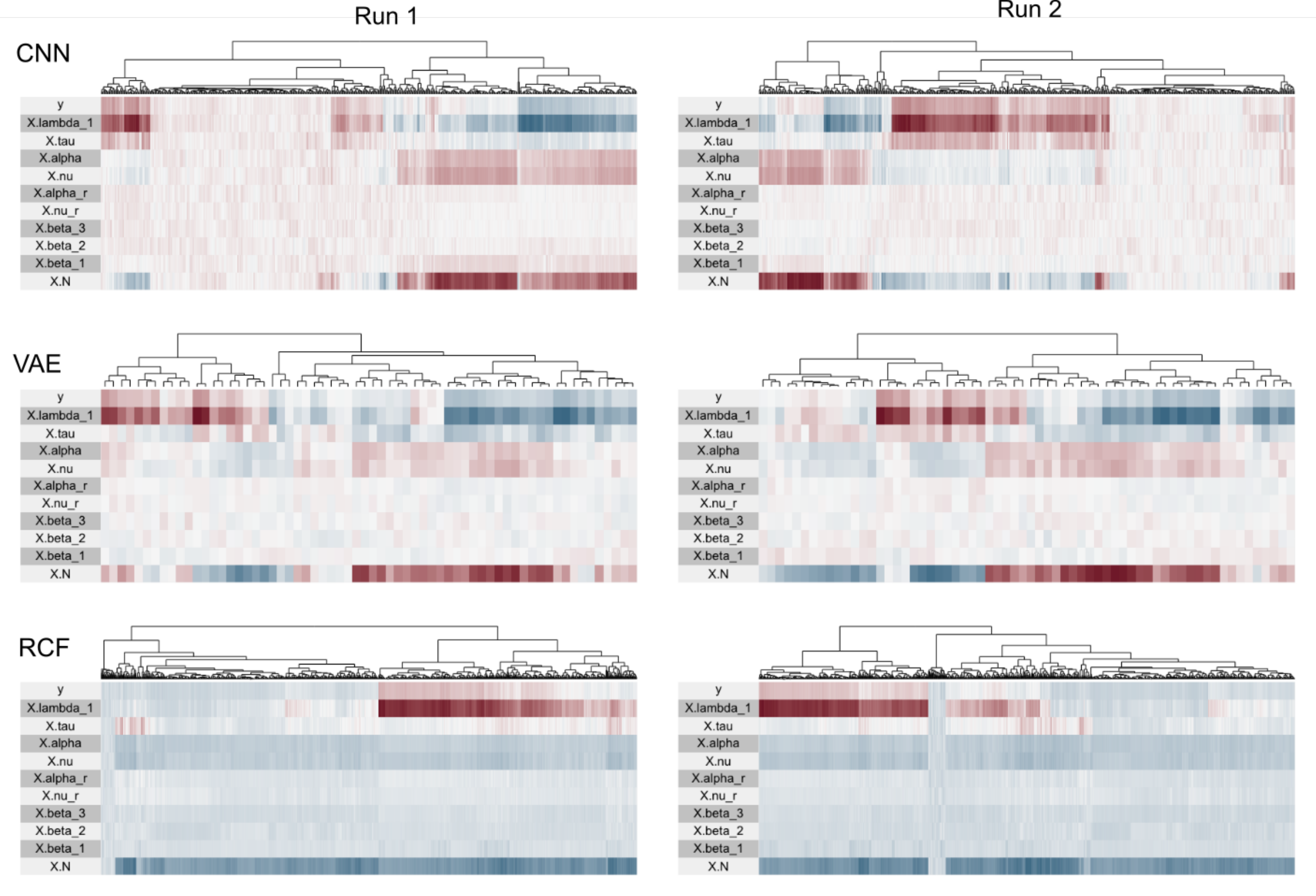}
  \caption{Each feature learning algorithm learns a distributed
    representation of the true latent features in the simulation. The two
    columns give results from two runs. Within each heatmap, rows correspond to
    the parameters in Table \ref{tab:sim_params}. Columns are activations of
    learned features; they have been reordered by a hierarchical clustering
    using the package \citep{barter2018superheat}. The color of a cell gives the
    correlation between the associated true and learned features. Blue and
    Burgundy encode positive and negative correlation, respectively. All learned
    features are associated with multiple true features at once.}
  \label{fig:distributed_hm}
\end{figure}

To compare the features learned by supervised and unsupervised approaches (G1),
we first directly visualize example learned features. Figure
\ref{fig:distributed_hm} shows the activations of learned features across 2000
images for two perturbed versions of the training data when $\absarg{I} = 0.9n$.
For the three algorithms, the learned features correspond to,
\begin{itemize}
\item CNN: Activations from the final hidden layer of neurons, used directly as input
  for the regression. There are a total of 512 nonnegative
  features\footnote{They are nonnegative because they follow an ReLU layer.}.
\item VAE: Spatially-pooled activations from the middle, encoding layer of the
  variational autoencoder. There are a total of 64 real-valued features.
\item RCF: The spatially-pooled activations corresponding to each of 1048 random
  convolutional features.
\end{itemize}

Our first observation is that, across algorithms, there is no simple
correspondence between learned and source features (i.e., parameters of the
underlying simulation). For example, it is not the case that one set of features
represents the number of cells $N$, and a different set maps to the roughness
$\nu_{\Lambda}$. Rather, there appear to be clusters of learned features, and
each cluster corresponds to a pattern of correlations across multiple source
features. For example, in Run 1 of the CNN, a cluster of learned features are
strongly negatively correlated with $y_{i}, \lambda_{i1}$, and $\tau$ and
positively correlated with $N_i$. We also find large subsets of features, across
all models, that are only weakly correlated with any source feature.

Next, note that certain source features are ``easier'' to represent than others,
in the sense that more of the learned features are strongly correlated with
them. Many features are correlated with $N_{i}$, the total number of cells in
the image, and $\lambda_{i1}$, the size of the cells from Process 1. Depending
on the model, the bandwidth $\alpha_{ir}$, roughness $\nu_{ir}$, and prevalence
$\beta_{ik}$ parameters are either only weakly or not at all correlated with the
learned features. Interestingly, the convolutional network learns to represent
$\beta_{1}$ well, but not $\beta_{2}$ or $\beta_{3}$ -- this is consistent with
the fact that only $\beta_{1}$ influences the response $y_{i}$. Even when
features learn to detect variation in $\alpha_{ir}$ and $\nu_{ir}$, they cannot
disambiguate between these two parameters.

Finally, consider differences between feature learning algorithms. The CNN and
VAE features tend to be more clustered, with strong correlation across several
source features. In contrast, the RCF features show more gradual shifts in
correlation strength. They also show relatively little variation in correlation
strength across features other than $\lambda_{i1}$ and $N_{i}$.

Note that the features do not naturally map onto one another from one run to the
next. This is not obvious from Figure \ref{fig:distributed_hm}, but a zoomed in
version in Supplementary Figure \ref{fig:distributed_hm_subset} showing only the
first 15 features per run makes this clear.

\subsubsection{Feature Learning vs. Inference}

\begin{figure}
  \centering
  \includegraphics[width=0.85\textwidth]{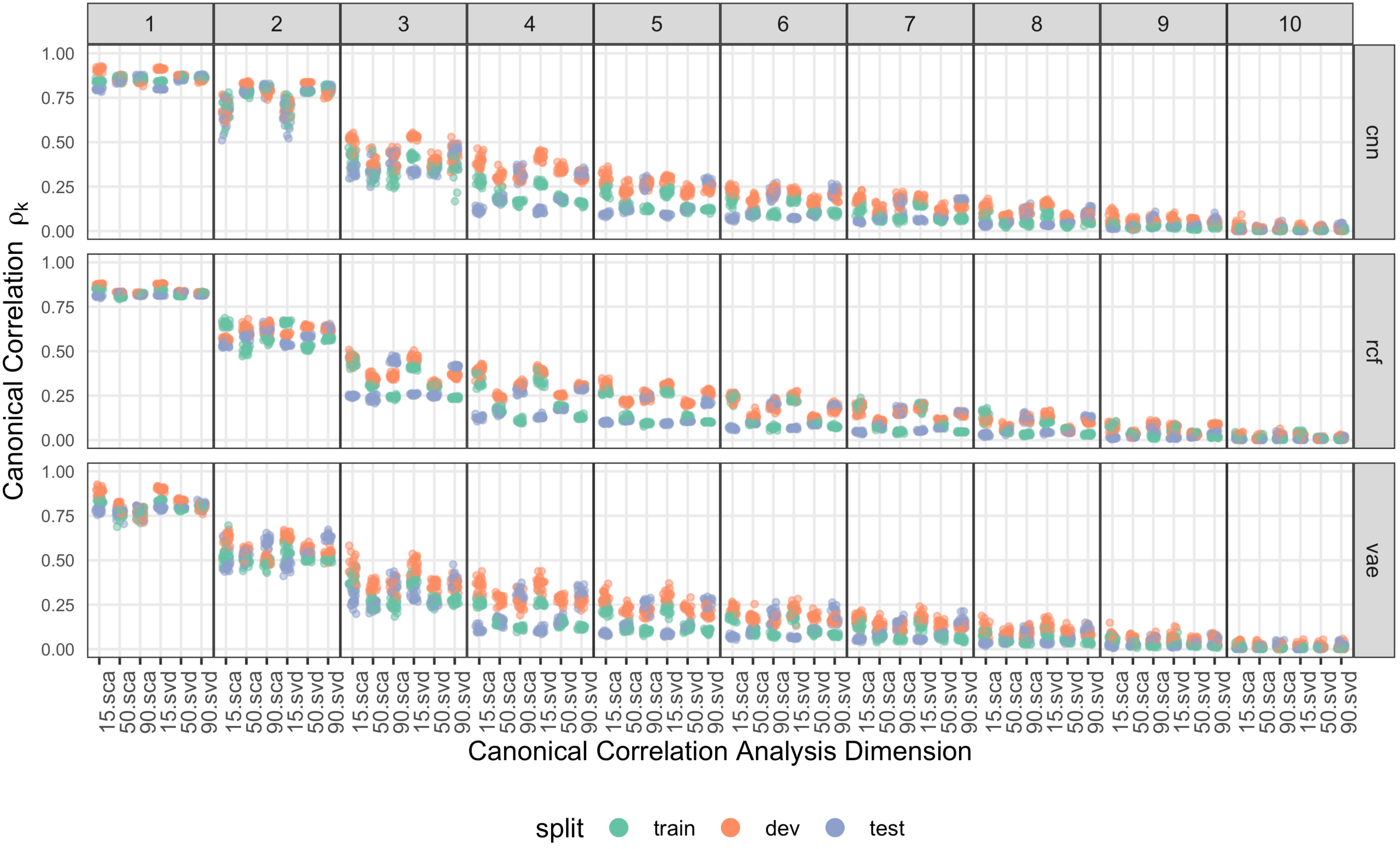}
  \caption{The top canonical correlations between the sources features and the
    aligned $\protect\underaccent{\bar}{\mathbf{Z}}_{b}$. Horizontal rows
    correspond to CNN, RCF, and VAE algorithms. Columns give the CCA dimension.
    Within each display, the canonical correlations across all bootstraps is
    shown. The $x$-axis within each cell gives the amount of data used for
    feature learning and the dimensionality reduction method used. Canonical
    correlations are computed separately for samples within training,
    development, and test splits, indicated by each point's color.}
  \label{fig:cca_summary}
\end{figure}

To consider the trade-offs between feature learning and inferential quality (G2)
and dimensionality reduction strategy (G3), Figure \ref{fig:cca_summary}
displays the top canonical correlations between learned features
$\underaccent{\bar}{\mathbf{Z}}_{b}$ and the original source features, across
bootstrap replicates and algorithms. Note that we calculate these scores
separately for training, development, and testing splits. The training and
development splits are subsets of $I$. Training samples were used in the
optimization for each feature learner; development samples were used to choose
hyperparameters of the optimizer. They are shown separate from the test samples
$I^{C}$ to make it possible to recognize potential overfitting in the feature
learning process.

Even after the initial dimensionality reduction, the CCA canonical correlations
decay quickly. The dimensionality reduction method used does not have much
effect at this stage. In general, the fraction of data belonging to $I$ also
matters little; however, there is an exception in the CNN features. Here, the
features learned when $\absarg{I}$ is only 15\% of the data have noticeably
lower canonical correlation in the second dimension. Note also that, from the
point of view of these canonical correlations, the RCF and VAE have comparable
feature learning behaviors.

The feature learning algorithms do not seem to overfit the simulation data. If
they did, then the canonical correlations on the training and development data
would be larger than those on the test data. That said, there are noticeable
differences between the data splits, and this effect will be visible in later
figures as well. We hypothesize that the differences are due to the alignment
process. During the Procrustes rotation, one matrix
$\mathbf{Z}_{b}\left[I^{C}\right]$ may get ``lucky'' and learn an alignment with
axes that better reflect variation in the source features. In this way, even
though the test data may not have been used to learn features, they may have
higher correlation with the source features than the original
$\mathbf{Z}_{b}\left[I\right]$.

\begin{figure}
  \centering
  \includegraphics[width=0.8\textwidth]{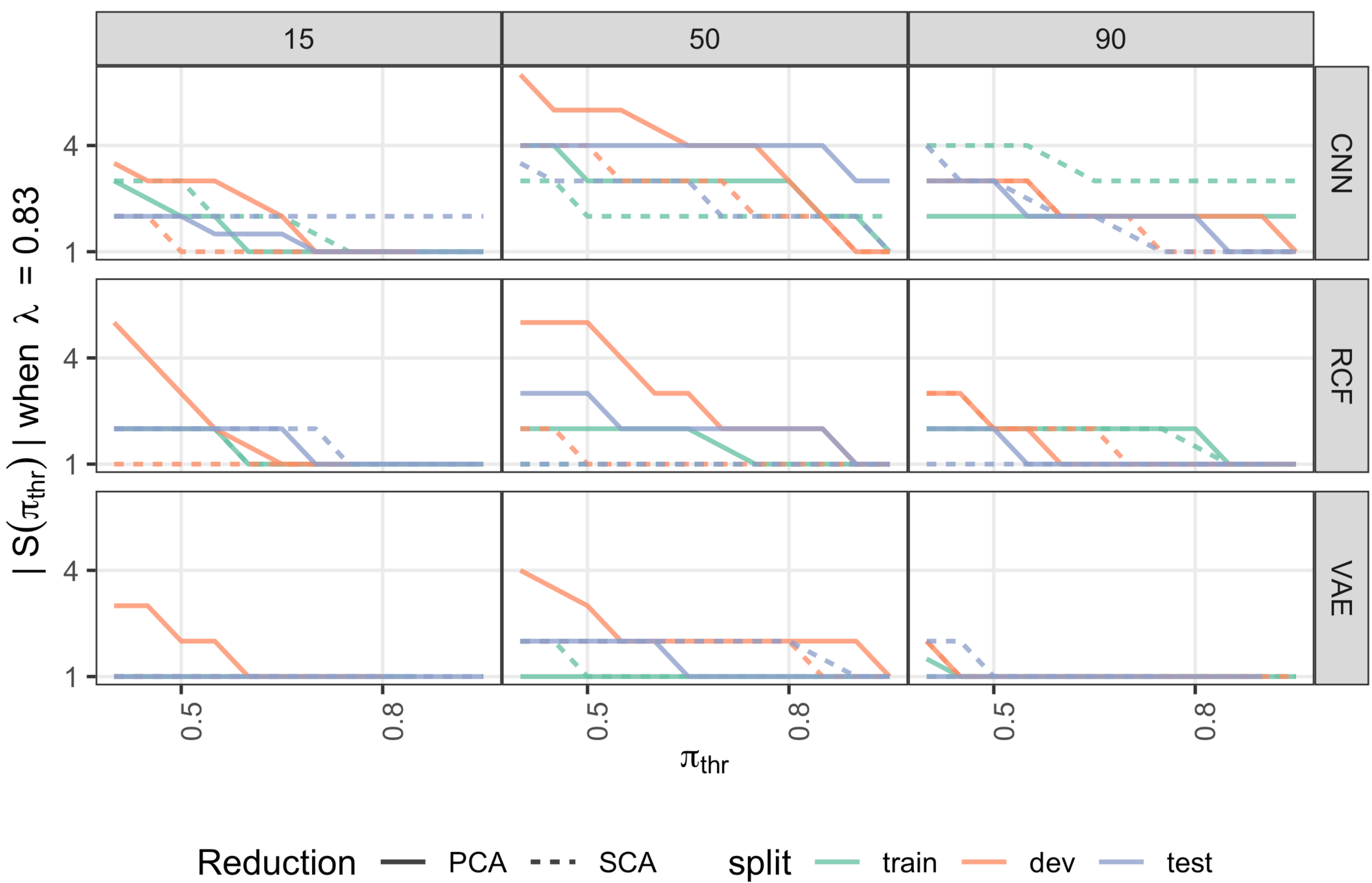}
  \caption{The median number of features selected by stability selection across
    bootstrap runs $\protect\underaccent{\bar}{\mathbf{Z}}_{b}$, viewed as a
    function of $\pi_{thr}$. Rows and columns correspond to models and fraction
    of the data used for training, respectively. Stability selection is run
    within training, development, and test splits (color) and using two
    dimensionality reduction procedures during alignment (solid vs. dashed
    lines).}
  \label{fig:selection_summary}
\end{figure}

To shed further light on G2 and G3, Figure \ref{fig:selection_summary} shows
the median number of features selected by stability selection
$\absarg{S\left(\lambda, \pi_{thr}\right)}$ across training sample sizes. The
median is taken across all bootstrap iterations. We fix a regularization
strength $\lambda$ where small perturbations in $\pi_{thr}$ lead to large
changes in the number of selected features. We have run stability selection for
$250$ replicates, all restricted to either training, development, or test data,
as indicated by the color of each line.

All the curves decrease because increasing the stringency $\pi_{thr}$ leads to
fewer selected features. We find that the features learned by the CNN are more
frequently selected. This is expected, considering that the CNN features are
learned in a supervised fashion. More surprising is the fact that the RCF
features are more frequently selected than the VAE features, suggesting that the
simple baseline might already provide features useful for interpretation, giving
an affirmative answer to G4.

Finally, it appears that $\absarg{S\left(\lambda, \pi_{thr}\right)}$ is largest
when using 50\% of the data for feature learning. Though the features learned
using 90\% of the data may be higher quality, it is not possible to select them
as frequently, because stability selection will have low power when it can only
subsample from the 10\% of the data reserved for inference. This phenomenon
appears for features learned by both supervised and unsupervised methods. For
this reason, in the remainder of this study, we will focus on results obtained
using a 50\% split between learning and inference, i.e., $\absarg{I} =
\absarg{I^{C}}$, though results using different splits are available in the
appendix.

\subsubsection{Stability visualization}

\begin{figure}
  \centering
  \includegraphics[width=\textwidth]{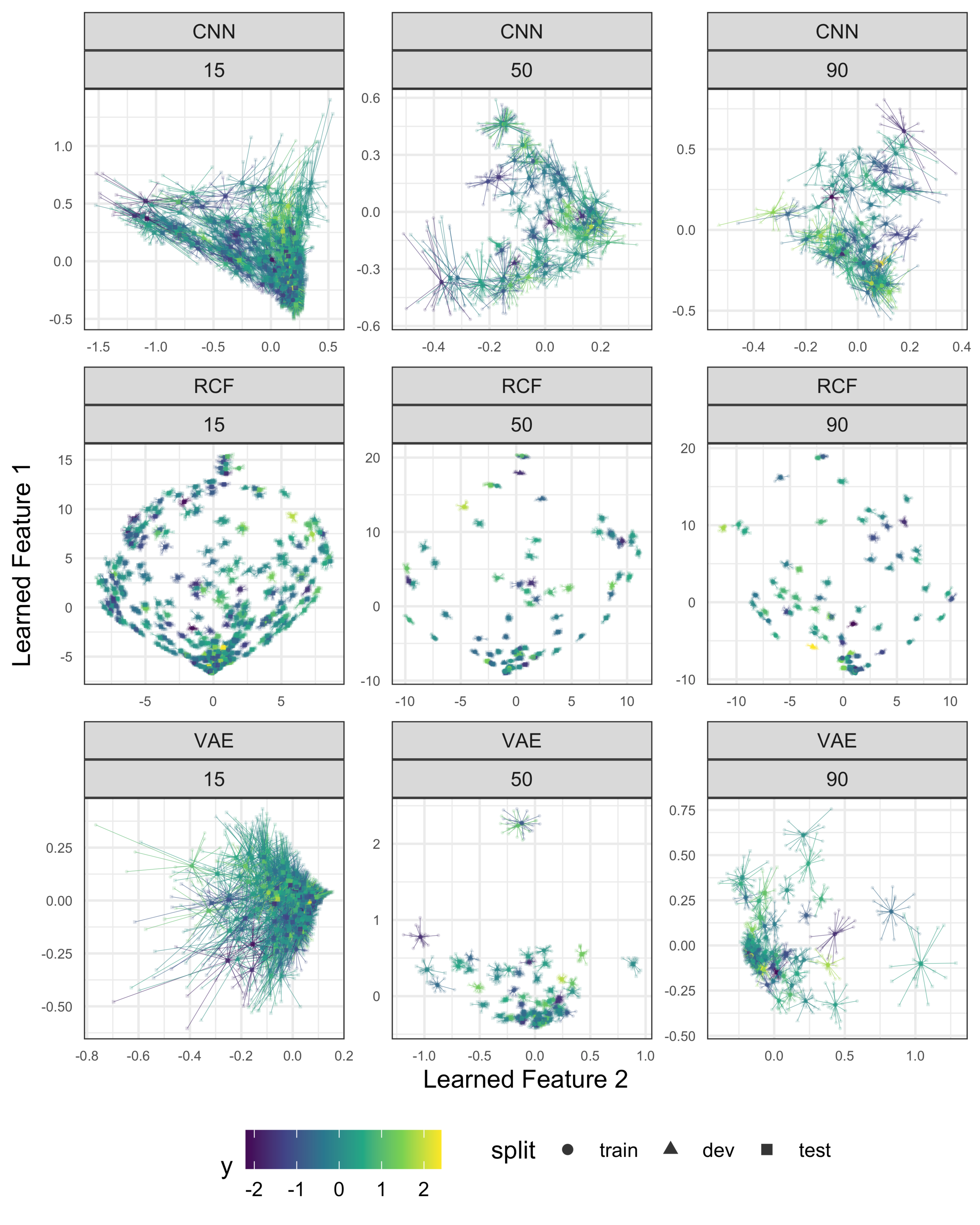}
  \caption{Example aligned features $\protect\underaccent{\bar}{\mathbf{Z}}_{b}$
    derived from several models with varying feature learning sample sizes
    $\absarg{I}$. For clarity, only a subset of samples is shown. For a version
    of this figure showing all samples, and collapsing each star to a point,
    refer to Supplementary Figure \ref{fig:sim_embeddings-full-pca-1-2}.}
  \label{fig:sim_embeddings-pca-1-2}
\end{figure}

\begin{figure}
  \centering
    \makebox[\textwidth][c]{\includegraphics[width=1.3\textwidth]{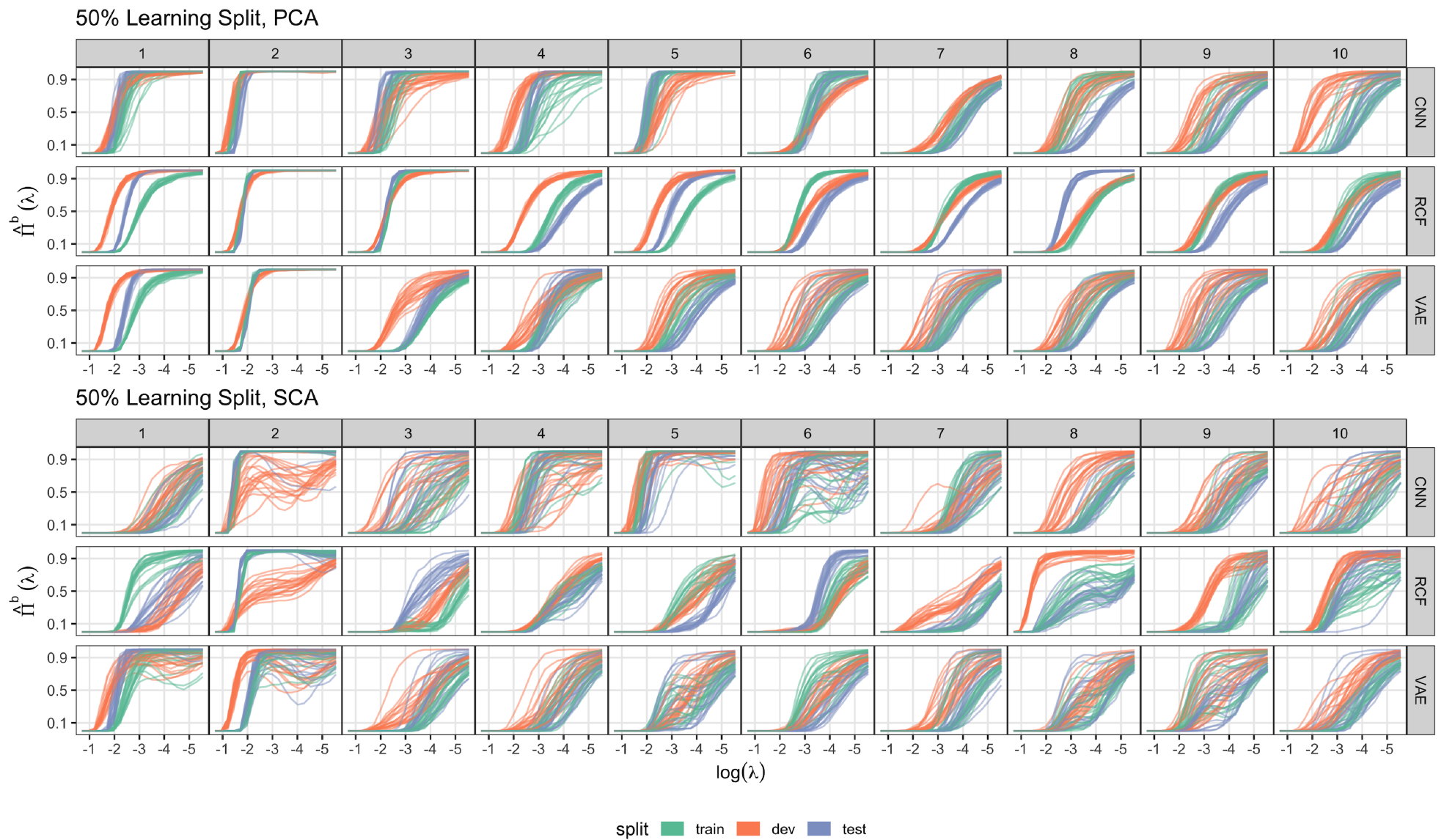}}
  \caption{Stability selection paths $\hat{\Pi}_{k}^{b}\left(\lambda\right)$ for
    10 learned features, across models and alignment strategies. Top and bottom
    groups use PCA and SCA-reduced features, respectively, and each row
    corresponds to a feature learning algorithm. Each column corresponds to one
    learned feature dimension. Each curve gives
    $\hat{\Pi}_{k}^{b}\left(\lambda\right)$ for one bootstrap replicate $b$
    restricted to samples from either the training, development, or test set.
    The $x$-axis is the regularization strength $\log \lambda$ and the $y$-axis
    is the probability of selecting that feature at the given regularization
    strength.}
  \label{fig:selection_paths-5-2}
\end{figure}

Figures \ref{fig:sim_embeddings-pca-1-2} and \ref{fig:selection_paths-5-2}
summarize the results of Algorithms \ref{alg:fss} and \ref{alg:selection}
applied to several feature learners. Within a single panel, each star glyph
corresponds to a single sample. The arms of the glyph link the bootstrap
replicates for that given sample: $\protect z_{i}^{1}, \dots, z_{i}^{20}$. Each
glyph is shaded by the true value of $y_{i}$ for that sample. Only the first two
aligned feature dimensions are shown.

\begin{figure}
  \centering
  \includegraphics[width=\textwidth]{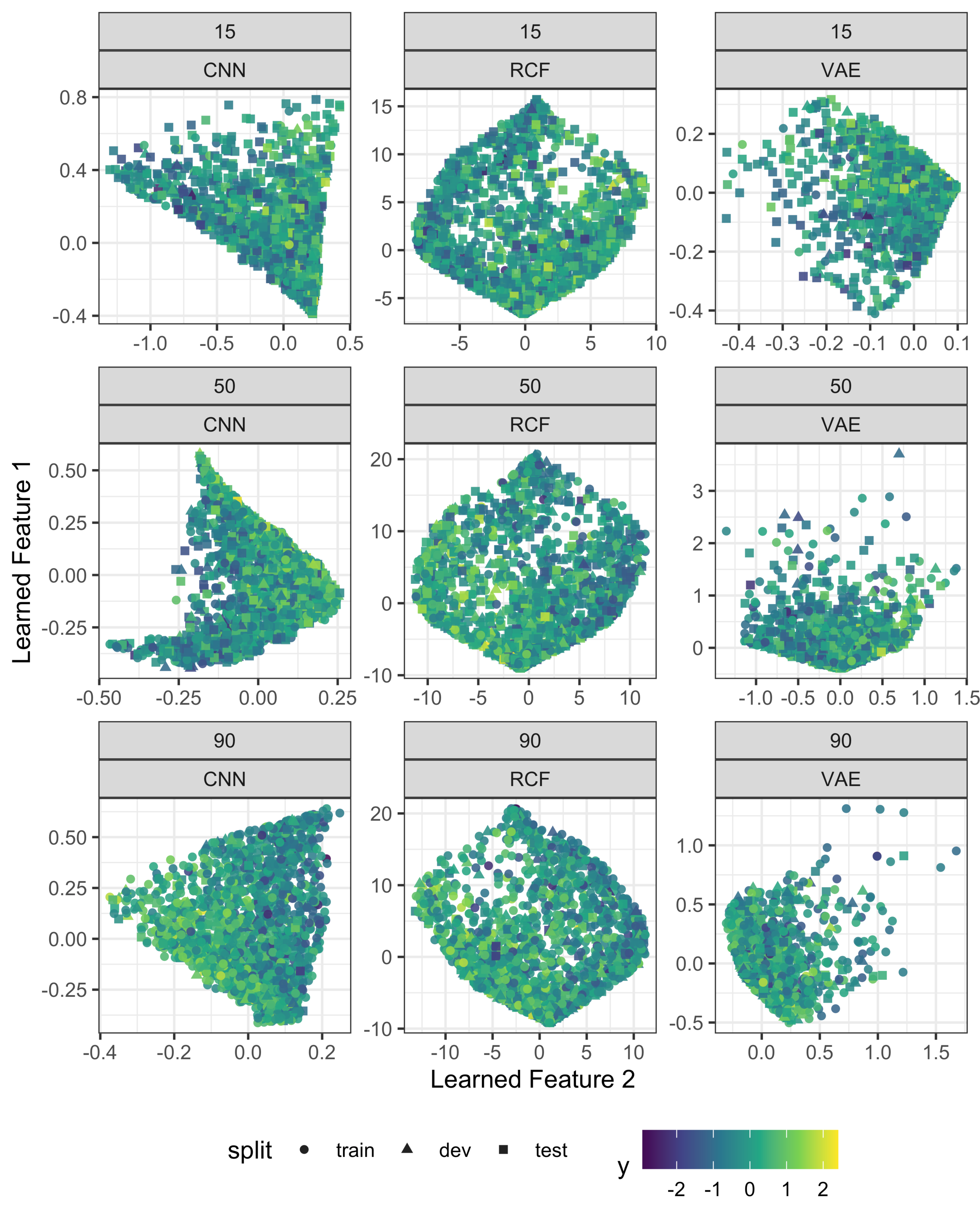}
  \caption{The analogous figure to Figure \ref{fig:sim_embeddings-pca-1-2}, but
    keeping only the center of each bootstrap alignment, and showing the
    embeddings for all images. This view makes the association with the response
    $y$ clearer, especially for larger $\absarg{I}$.}
  \label{fig:sim_embeddings-full-pca-1-2}
\end{figure}

Both the model and reduction strategy used influence the stability of the
learned features. Based on the relative sizes of the glyphs, the CNN features
are least stable, those from the RCF are most stable, and the VAE features are
intermediate. Separate regions of the learned feature space appear more stable
than others. Features learned with 15\% of the data are highly unstable and show
little association with the response. For larger $\absarg{I}$'s, the learned
features are more stable. Supplementary Figure
\ref{fig:sim_embeddings-full-pca-1-2} shows that in this case, there are also
stronger associations with $\mathbf{y}$.

Figure \ref{fig:selection_paths-5-2} displays stability selection paths
$\Pi_{k}^{b}\left(\lambda\right)$ for combinations of feature learning and
dimensionality reduction procedures. Like in Figure \ref{fig:cca_summary}, we
find that the largest variation is between the data splits, but that
overfitting does not appear to be a problem. Indeed, for several features, the
development and test splits have higher selection probabilities.

Revisiting the effect of dimensionality reduction approach on selection (G3), we
find the first major differences between dimensionality reduction strategies.
The selection paths are not monotone for SCA, and they also vary substantially
across runs. This is most likely a result of the fact that the matrix $B$ in the
decomposition $X = Z B Y^{T}$ need not be diagonal, which correlates the
resulting coordinates. Further, when using the PCA, the top features are also
the most selected ones; this is not the always the case with SCA.

Next, compare the unsupervised, supervised, and accelerated methods (G1, G4),
from the selection perspective. Between algorithms, we find that the selection
paths for RCF features form a tighter band across all bootstrap samples, another
mark in its favor. For all algorithms, this band seems to widen for later
feature dimensions; the selection probabilities also generally rise more
gradually. For Feature 3 onwards, the selection paths for the VAE rise more
gradually than the corresponding paths for either the CNN or RCF dimensions,
suggesting a lack of association with $\mathbf{y}$.

\section{Data Analysis}
\label{sec:dataset}

In this section, we conduct a feature stability analysis on the spatial
proteomics dataset\footnote{The data are
  \href{https://www.angelolab.com/mibi-data}{publicly available}. See also the
  appendix for preprocessed versions.} reported in the study
\citep{keren2018structured}, where the authors found a relationship between the
spatial organization of Triple Negative Breast Cancer (TNBC) tissues and disease
progression. In a classical proteomics study, the expression levels for a set of
proteins is measured for a collection of cells, but the cell locations are
unknown. In contrast, these data provide for each patient (1) an image
delineating cell boundaries and (2) the protein expression levels associated
with each cell in the images.

We will work only with the spatial cell delineations, not the protein expression
levels. This allows us to study the mechanics of feature learning within the
images without having to worry about linking the expression and image data,
which is in itself a complex integration problem. Though this means we lose some
scientific depth, we gain substantial implementation simplicity, and the
analysis serves as a clear illustration. Our complete data are 41 $2048 \times
2048$-dimensional images, each taken from a separate patient. We associate each
pixel with one of 7 categories of tumor and immune cell types.

\begin{figure}
  \centering
  \includegraphics[width=\textwidth]{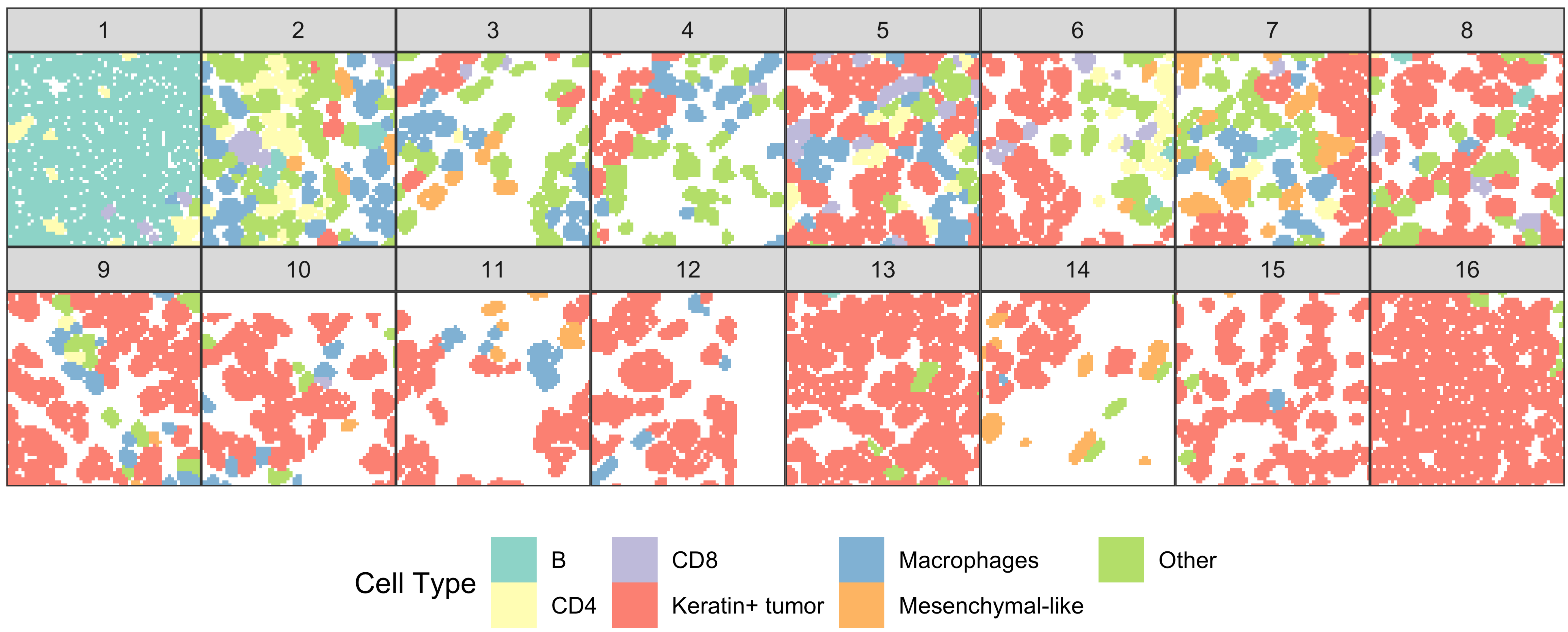}
  \caption{Example patches from the TNBC data. Panels are ordered by $y_i$, the
    (log) fraction of cells they contain that belong to the tumor. The goal of
    the prediction algorithm in this example is to correctly place patches from
    new patients along this gradient, after having observed many patches and
    their corresponding $y_i$'s.}
  \label{fig:example_cells}
\end{figure}

To setup a prediction problem, we first split each image into $512 \times 512 \times 7$
patches. These patches are our $x_{i}$. Patches from 32 of the patients are
reserved form feature learning. Four among these 32 are used as a development
split, to tune parameters of the feature learning algorithms. As a response
variable, we use $y_{i} = \log\left(\frac{\#\{\text{Tumor cells in
  }x_{i}\}}{\#\{\text{Immune cells in }x_i\}}\right)$. Example cell patches are
shown in Figure \ref{fig:example_cells}.

\begin{figure}
  \centering
  \includegraphics[width=0.8\textwidth]{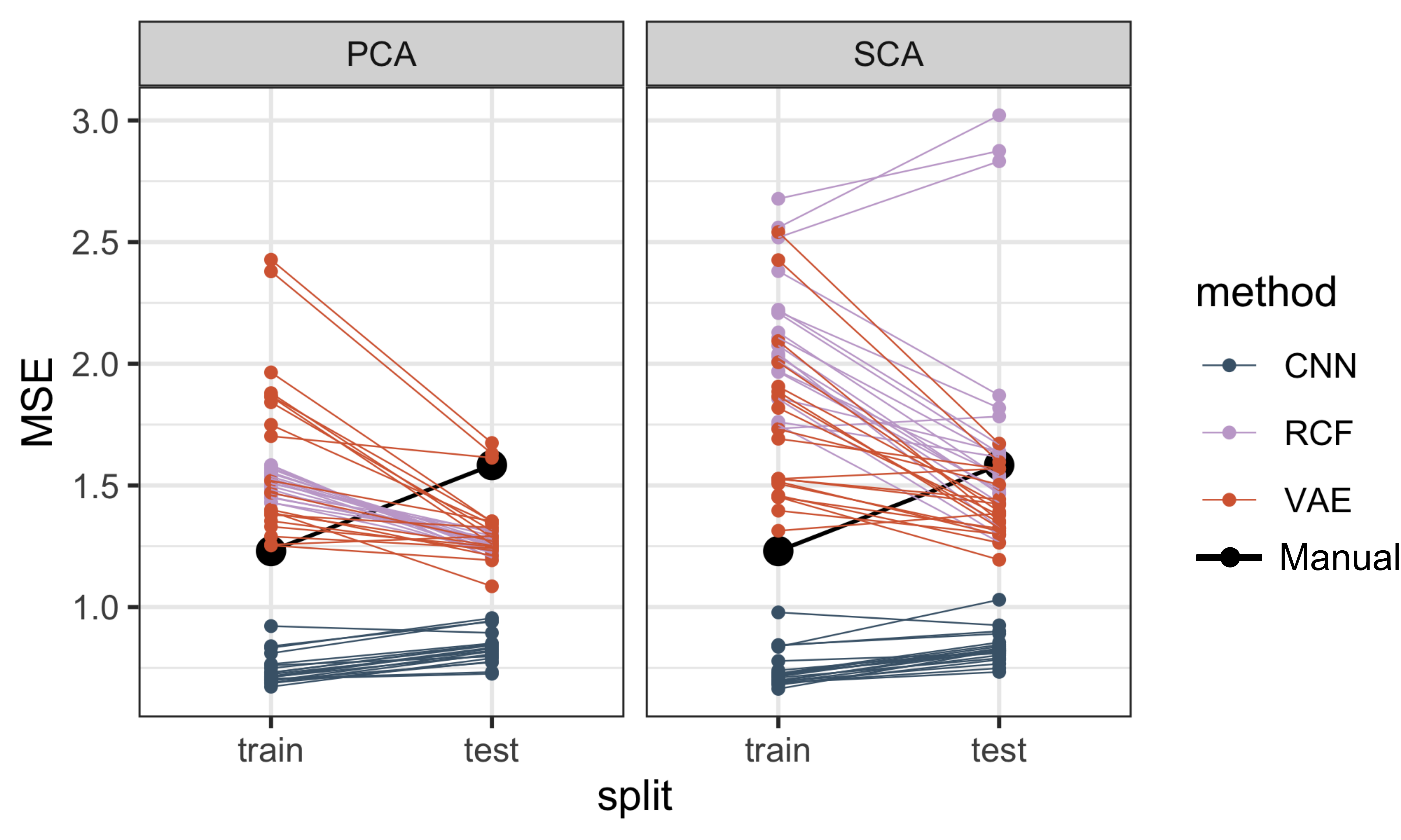}
  \caption{Relative performance of feature learning strategies on the TNBC data.
    Linked points come from the same bootstrap replicate. The manual features
    refers to the regression model that simply uses the raw pixel counts for
    each of the cell types. Predictions from the development split are omitted;
    this split has few patches, and the estimated MSE's have high variance.}
  \label{fig:tnbc_baseline}
\end{figure}

As a baseline, we compare against a ridge regression with pixelwise composition
features. Specifically, we train a model with $\mathbf{y}$ as a response and the
average number of pixels belonging to each of the cell-type categories as a
(length 7) feature vector. This helps us to determine whether the model has
learned more interesting features useful for counting cells, like cell size and
boundaries, rather than simply averaging across pixel values. Indeed, Figure
\ref{fig:tnbc_baseline} makes clear that, with the exception of the RCF-SCA
combination, all feature learning - dimensionality reduction combinations
outperform this manual baseline.

\begin{figure}
  \centering
    \makebox[\textwidth][c]{\includegraphics[width=1.3\textwidth]{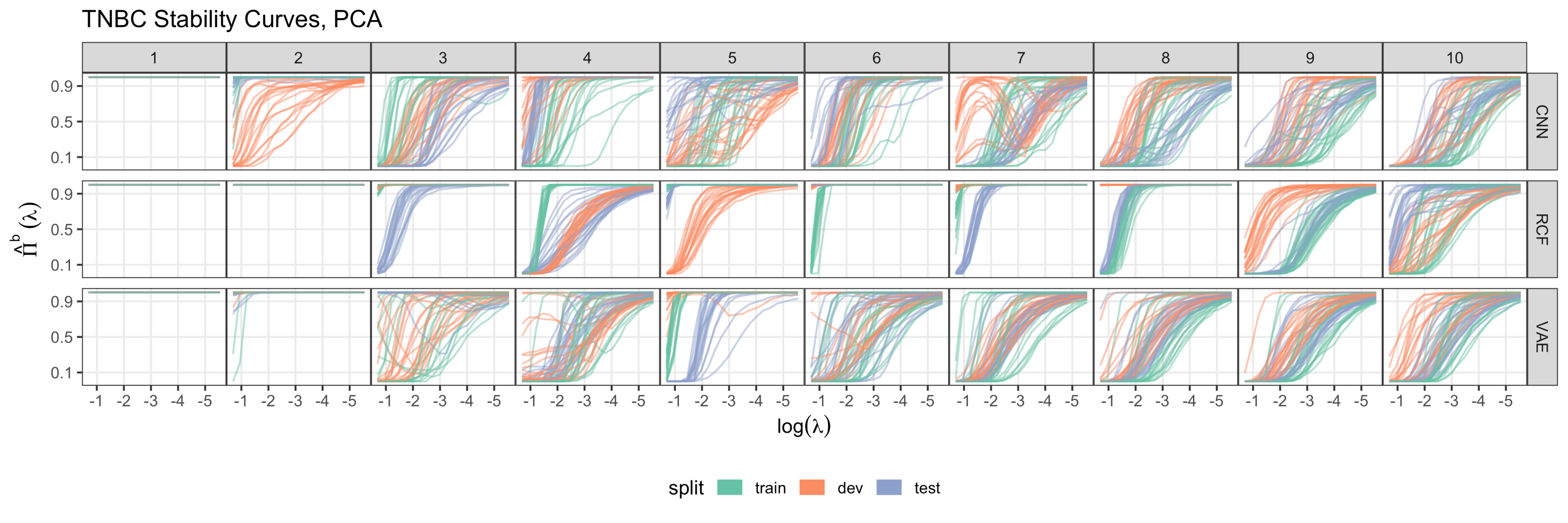}}
  \caption{Selection curves for features learned from the TNBC data. 50\% of
    the data were used for feature learning and PCA was used for
    dimensionality reduction. Each curve is read as in Figure
    \ref{fig:selection_paths-5-2}. The analogous paths using SCA alignment are
    given in Supplementary Figure \ref{fig:tnbc_selection_paths-1}.}
  \label{fig:tnbc_selection_paths-2}
\end{figure}

Stability curves associated with the learned features from the CNN, RCF, and VAE
models are shown in Figure \ref{fig:tnbc_selection_paths-2}. Interestingly,
across all algorithms, the top aligned features are not necessarily those with
the highest selection probabilities. For example, CNN Feature 4 has higher
selections than Feature 3, RCF Feature 6 has higher selection than Feature 4,
and VAE Feature 5 is more frequently selected than Feature 3. In contrast to the
simulation, the CNN and VAE do not have substantial differences in their
selection curves. Further, the RCF features seem to have high selection
probabilities across more dimensions than either the CNN or VAE. This suggests
that the most salient features in $x_i$ are also relevant for predicting $y_i$,
and that supervision is not as critical in this problem as it was in the
simulation.

\begin{figure}
  \centering
  \includegraphics[width=\textwidth]{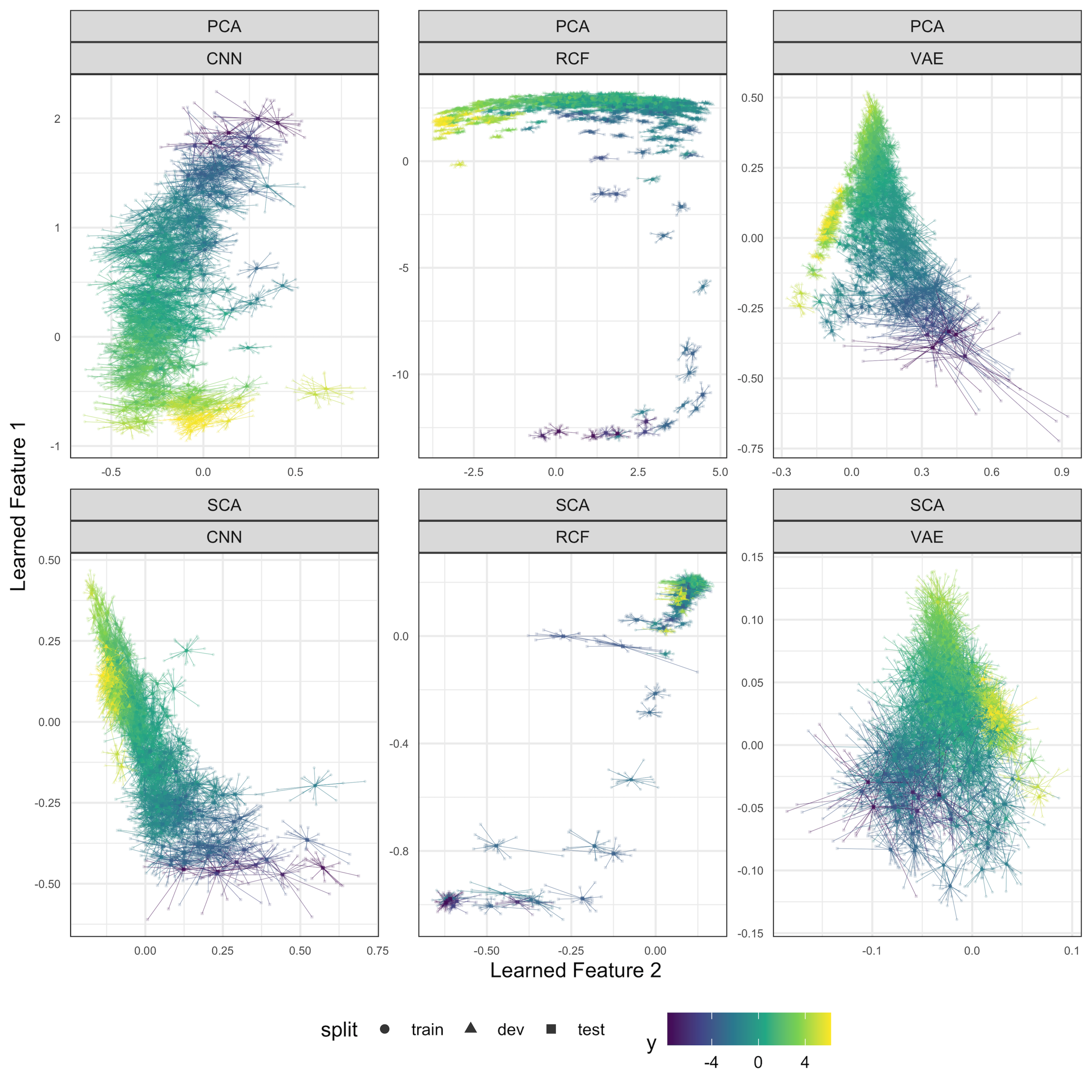}
  \caption{Representation of samples according to some of the important learned
    feature dimensions. Glyphs are read as in Figure
    \ref{fig:sim_embeddings-pca-1-2}. To interpret these features, example
    patches are arranged according to these coordinates in Figure
    \ref{fig:tnbc_imagegrid-PCA-1-2} and Supplementary Figure
    \ref{fig:tnbc_imagegrid-SCA-1-2}. Only a subsample of points are shown; the
    full dataset with stars collapsed to points is shown in Figure
    \ref{fig:tnbc_embeddings-full-both-1-2}.}
  \label{fig:tnbc_embeddings-1-2}
\end{figure}

Example aligned coordinates are given in Figure
\ref{fig:tnbc_embeddings-1-2}. Consistent with the conclusion, we find that
the association with the response is clearly visible with respect to the learned
feature dimensions, even when using unsupervised algorithms. The changes in the
sizes of glyphs across regions of the learned feature space are especially
pronounced in this application. For example, in the VAE, the representations of
samples with higher tumor-to-immune ratio $y_i$ are much more stable than those
with low ratio.

\begin{figure}
  \centering
  \makebox[\textwidth][c]{\includegraphics[width=1.2\textwidth]{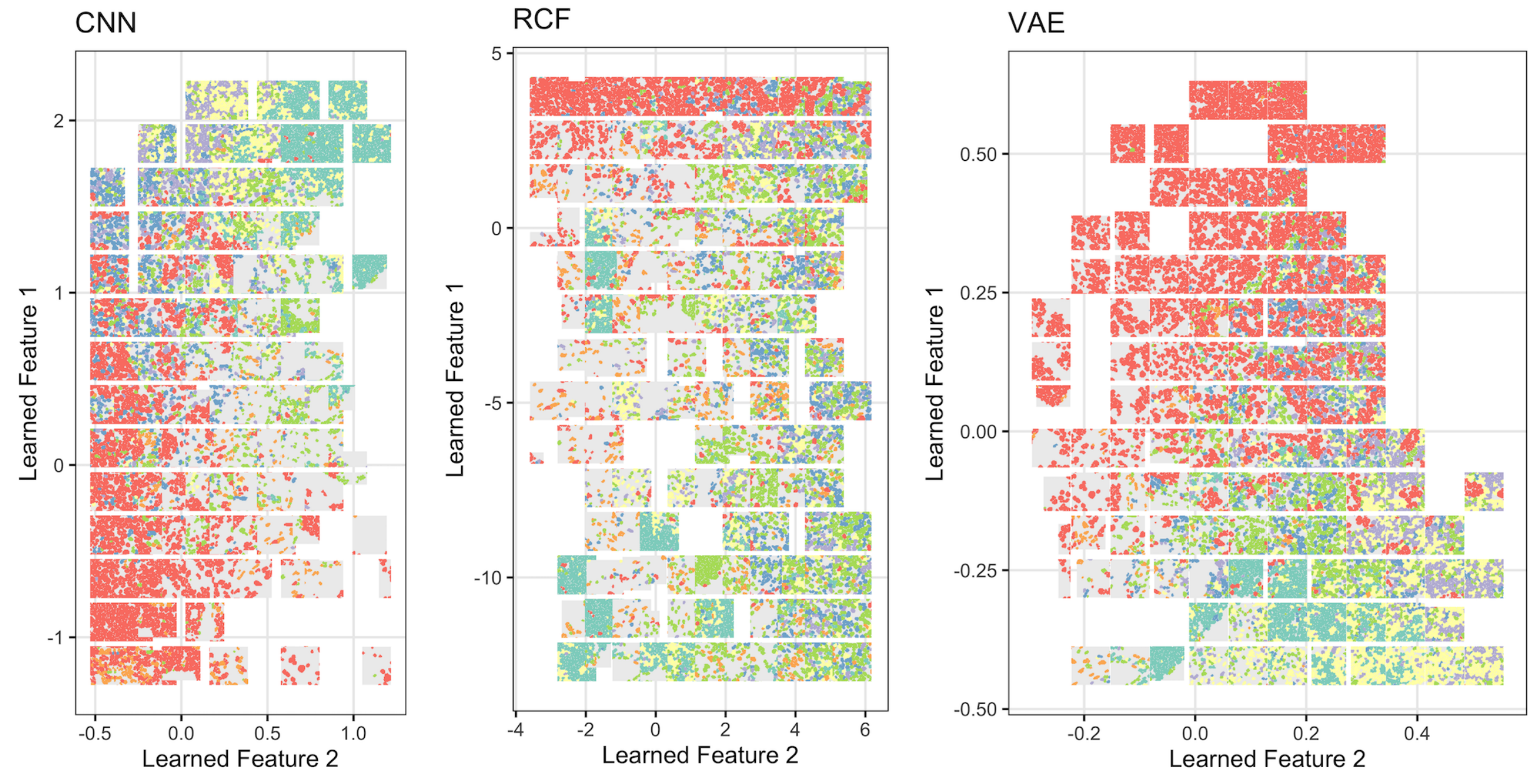}}
  \caption{A version of the PCA panels of Figure \ref{fig:tnbc_embeddings-1-2}
    that overlays representative samples across the learned feature space. Cells
    are color coded as in Figure \ref{fig:example_cells}. Note that the overall
    shape of the region in which images are displayed mirrors the shape of the
    cloud of points in Figure \ref{fig:tnbc_embeddings-1-2}.}
  \label{fig:tnbc_imagegrid-PCA-1-2}
\end{figure}

There is no consensus on how to best interpret automatically learned features
\citep{doshi2017towards}. Nonetheless, we present one simple approach in Figure
\ref{fig:tnbc_imagegrid-PCA-1-2}, overlaying example patches onto aligned
coordinates. For example, in the CNN, the first dimension distinguishes between
the relative number of tumor and immune cells, while the second dimension
reflects the density of cells. In the RCF, the second dimension captures the
diversity of cell types, with more uniform samples on the left and more
heterogeneous ones on the right. The second dimension of the VAE seems related
to both cell density and number of cell types. The analogous display from an SCA
reduction is given in Supplementary Figure \ref{fig:tnbc_imagegrid-SCA-1-2}.

\section{Discussion}

This study has investigated the stability of machine-generated, rather than
hand-crafted, features. A better understanding of stability in this modern
regime has consequences for how these methods can be used in real-world
applications, especially those intended for scientific workflows.

Our results raise several questions for further study. It is natural to ask
to what extent similar behaviors are exhibited across other data domains, model
types, or training regimes. For example, it would not be unusual to represent
the cell data in our case study using a marked graph linking neighboring cells.
Do the features learned by a graph autoencoder have similar stability
properties? In other domains, we may ask whether our methods can be adapted to
text or audio data.

Further, there are questions that may guide us towards better instantiations of
Algorithms \ref{alg:features} through \ref{alg:selection}. While we have relied
on the bootstrap, our notation encompasses more general perturbations
$\mathcal{P}$. For example, how might learned features change when discarding
nonrandomly chosen subsets of training data? Perhaps learning features based on
different temporal or spatial subsets could reveal a drift in the important
features over time or space. Alternatively, we could imagine perturbing the
model training procedure, using different hyperparameters. It would be of
interest to trace out the dependence of the learned features on the mechanics of
the feature learner.

Similarly, using a Procrustes rotation for $\mathcal{A}$ and stability selection
for $\mathcal{S}$ provides a reasonable point of departure, but more
sophisticated approaches are possible. For example, dimensionality reduction
could be optimized to support alignment; this could be accomplished using
multiple canonical correlation analysis. Combinations of features could also be
approximately matched across feature learners, using a form of optimal
transport, identifying sets of features that all activate on similar input
samples. This has been applied in a federated learning context, but not in the
study of stability \citep{wang2020federated}.

It is also possible to propose alignments based on more refined measures of
correlation \citep{josse2016measuring, azadkia2019simple}. We have found that
using an even split between feature learning and inference gives reasonable
results, but our results are admittedly coarse. Though we have concentrated on
stability selection, the procedure $\mathcal{S}$ could be any selective
inference procedure. Finally, our approaches to summarizing and displaying the
resulting representations may be improved through a more careful application of
interactive visualization principles.

More broadly, this study is situated in the body of work seeking to bridge the
rift between the two cultures \citep{breiman2001statistical,
  efron2020prediction}. Across the selective inference, conformal prediction,
and interpretability literatures, there is a growing understanding that the
statistics and machine learning communities could benefit by learning to speak
one another's languages \citep{angelopoulos2020uncertainty, ren2020knockoffs}.
Nonetheless, while the data sources and feature learning context we consider is
novel, our basic motivation is an old statistical idea that still rings true --
quoting from \citep{mosteller1977data},

\begin{quote}
One hallmark of the statistically conscious investigator is a firm belief that,
however the survey, experiment, or observational program actually turned out, it
could have turned out some somewhat differently.
\end{quote}

In order to accomplish this study, we have adapted tools from the
representational analysis, dimensionality reduction, and high-dimensional
inference communities. These tools give a window into the workings of modern
feature extraction techniques, helping us view commonplace algorithms in a new
way. In a sense, our intent is to do more than summarize data -- it is to
generate new data. In the same way that a standard error around a mean is a new
piece of information that supports more nuanced reasoning, we hope to see the
development of techniques that generate data characterizing the behavior of
modern feature learning algorithms.

\bibliographystyle{plainnat}
\bibliography{refs}

\section{Reproducibility}

Instructions to reproduce our simulations and data analysis example are
available in a \href{https://github.com/krisrs1128/learned_inference}{README} on
the study's github page. Training split creation, the regression baseline, and
feature learning can be reproduced in the ipython notebooks

\begin{itemize}
\item \texttt{tnbc\_splits.ipynb}: Define training and test splits for the TNBC
  dataset.
\item \texttt{tnbc\_baseline.ipynb}: Train a ridge regression baselien on the
  TNBC dataset.
\item \texttt{model\_training.ipynb}: Train either a CNN, RCF, or VAE feature
  learner.
\end{itemize}

MIBI-ToF data preparation, generation of simulation data, feature stability
analysis, and visualization of results are done within the rmarkdown documents,

\begin{itemize}
\item \texttt{generate.Rmd}: Simulate the LCGP cells dataset.
\item \texttt{stability.Rmd}: Perform feature alignment and stability selection.
\item \texttt{visualize\_features.Rmd}: Visualize the aligned and selected
  features output by \texttt{stability.Rmd}.
\item \texttt{summary\_plots.Rmd}: Compute the CCA and selection summaries
  reported in Figures \ref{fig:cca_summary} and \ref{fig:selection_summary}.
\end{itemize}

To support code reusability between experiments, two helper packages were
prepared,
\begin{itemize}
\item \href{https://github.com/krisrs1128/learned_inference/tree/master/stability}{\texttt{stability}}
\item \href{https://github.com/krisrs1128/learned_inference/tree/master/inference}{\texttt{inference}}
\end{itemize}

This packages can be installed by calling,
\begin{verbatim}
git clone https://github.com/krisrs1128/learned_inference.git
Rscript -e ``devtools::install(`learned_inference/inference')''
pip3 install learned_inference/stability
\end{verbatim}
from the terminal.

We have prepared a \href{https://hub.docker.com/r/krisrs1128/li}{docker image}
with all necessary software pre-installed. For example, to rerun our stability
analysis, the following commands may be used,

\begin{verbatim}
docker run -it krisrs1128/li:latest bash
git clone https://github.com/krisrs1128/learned_inference.git
source learned_inference/.env
# download relevant data
Rscript rmarkdown -e ``rmarkdown::render(`learned_inference/inference/vignettes/stability.Rmd')''
\end{verbatim}

Finally, we have released raw data and intermediate results from our analysis,
\begin{itemize}
\item
  \href{https://drive.google.com/file/d/1v_Ndux1Rmk2q1ul5Vv5srgI1JQ17Vx0n/view?usp=sharing}{sim\_data.tar.gz}:
  Our toy simulation dataset.
\item \href{https://drive.google.com/file/d/1KMG5yrty8UEPhrR0Y7hIZrtwWuP_y-cm/view?usp=sharing}{tnbc\_data.tar.gz}: The preprocessed MIBI-ToF data, with all
  patient's data split into 64 $\times$ 64 patches and with the associated
  splits and response value stored in a metadata file.
\item \href{https://drive.google.com/file/d/1QVmyqYQCe8C04rAyBQxXuZYuaiYleopx/view?usp=sharing}{simulation\_outputs.tar.gz}: All the models trained in our
  simulation experiments.
\item \href{https://drive.google.com/file/d/1DmzObBWCzVzDNZ1DxcUWgyIoZmC4H8gD/view?usp=sharing}{tnbc\_outputs.tar.gz}: All models trained in our illustration on
  the TNBC dataset.
\item \href{https://drive.google.com/file/d/1zJQOB2dSuy_1WYheJtUI8PZfu9PsyVGQ/view?usp=sharing}{simulation\_figure\_data.tar.gz}: The data written by
  `stability.Rmd` which was used to generate figures for our simulation.
\item \href{https://drive.google.com/file/d/1FaCrOysBlsNYgzul6iJiFcmgRFl2OaSE/view?usp=sharing}{tnbc\_figure\_data.tar.gz}: The data written by `stability.Rmd`
  which was used to generate figures for our data illustration.
\item \href{https://drive.google.com/file/d/1c-ZPs9RbzkY9FzUnCCVJc8syt5dpj-4Y/view?usp=sharing}{tnbc\_raw.tar.gz}:
  The original MIBI-ToF tiffs, before splitting into patches.
\end{itemize}

\section{Supplementary Figures}

\begin{figure}
\includegraphics[width=\textwidth]{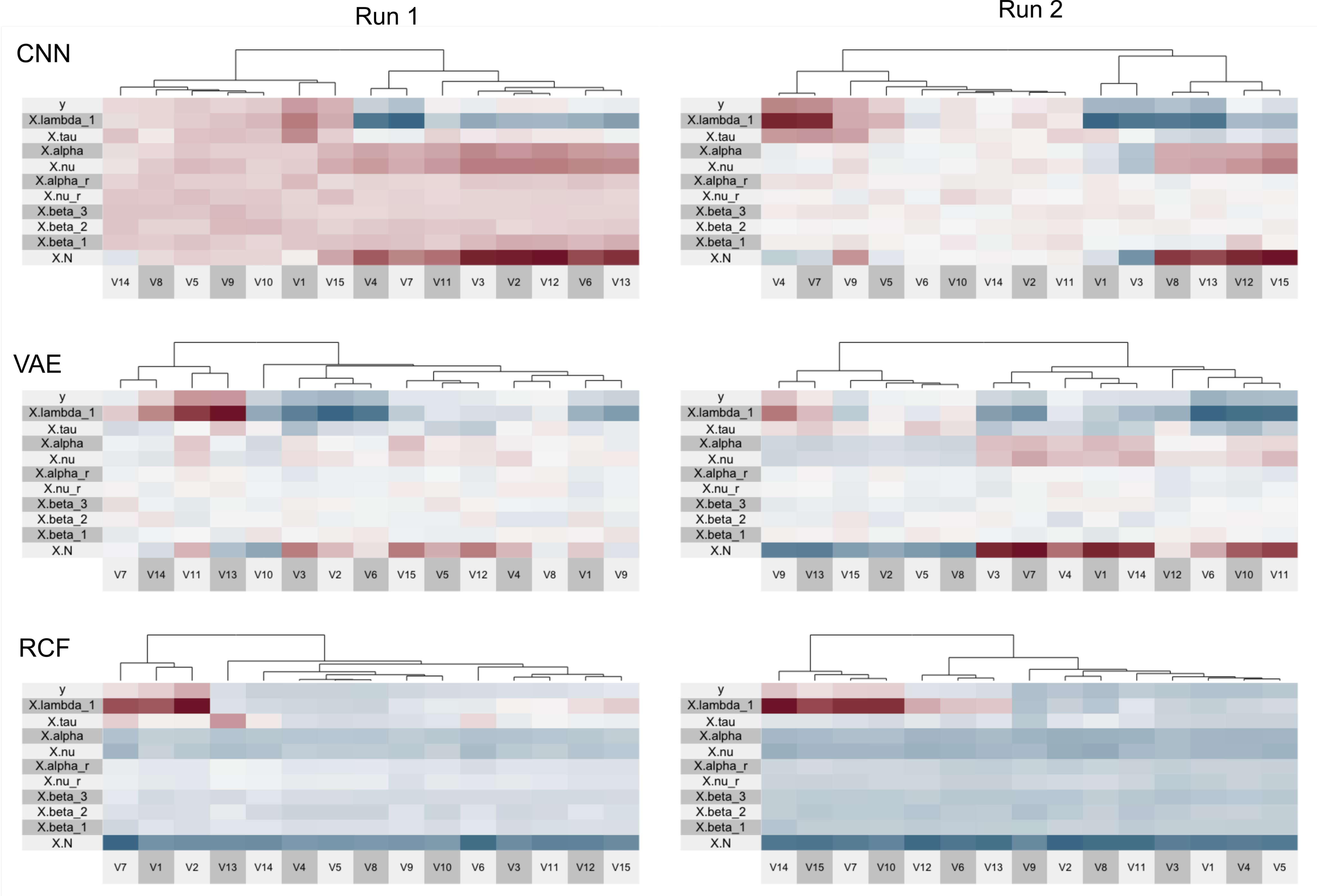}
\caption{A zoomed version of Figure \ref{fig:distributed_hm}, showing that the
  features cannot be directly mapped onto one another, from one run to the
  next.}
\label{fig:distributed_hm_subset}
\end{figure}

\begin{figure}
\includegraphics[width=\textwidth]{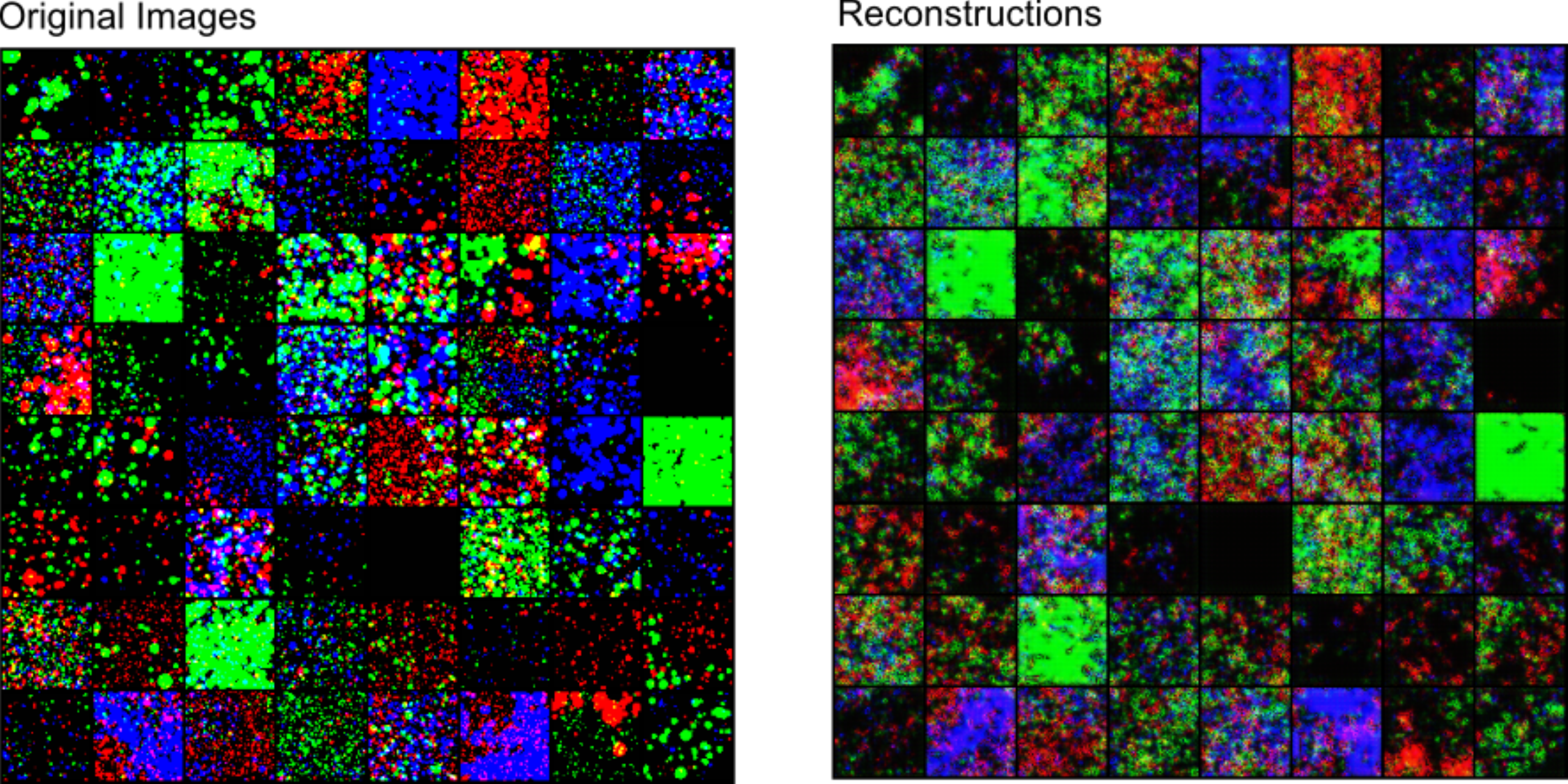}
\caption{Example reconstructions from the VAE model applied to the simulation
  data. Original image patches are shown on the left, corresponding
  reconstructions are given on the right. Though fine-grained details are missed
  by our version of the VAE model, most of the key global features of each patch
  seem accurately reflected.}
\label{fig:reconstructions}
\end{figure}

\begin{figure}
  \centering
  \includegraphics[width=\textwidth]{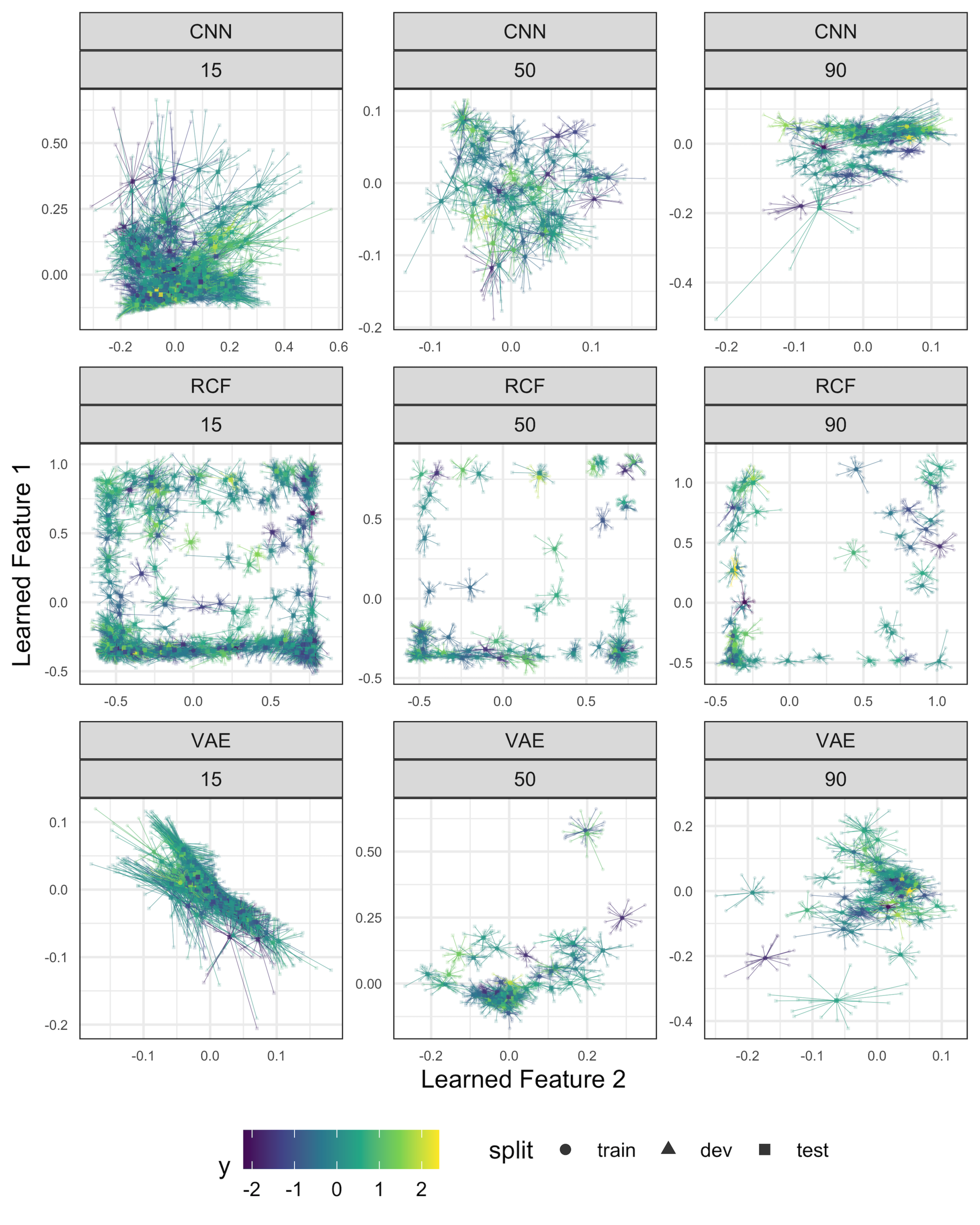}
  \caption{The analogous figure to Figure \ref{fig:sim_embeddings-pca-1-2}, but
    using SCA dimensionality reduction.}
  \label{fig:sim_embeddings--sca-1-2}
\end{figure}

\begin{figure}
  \centering
  \includegraphics[width=\textwidth]{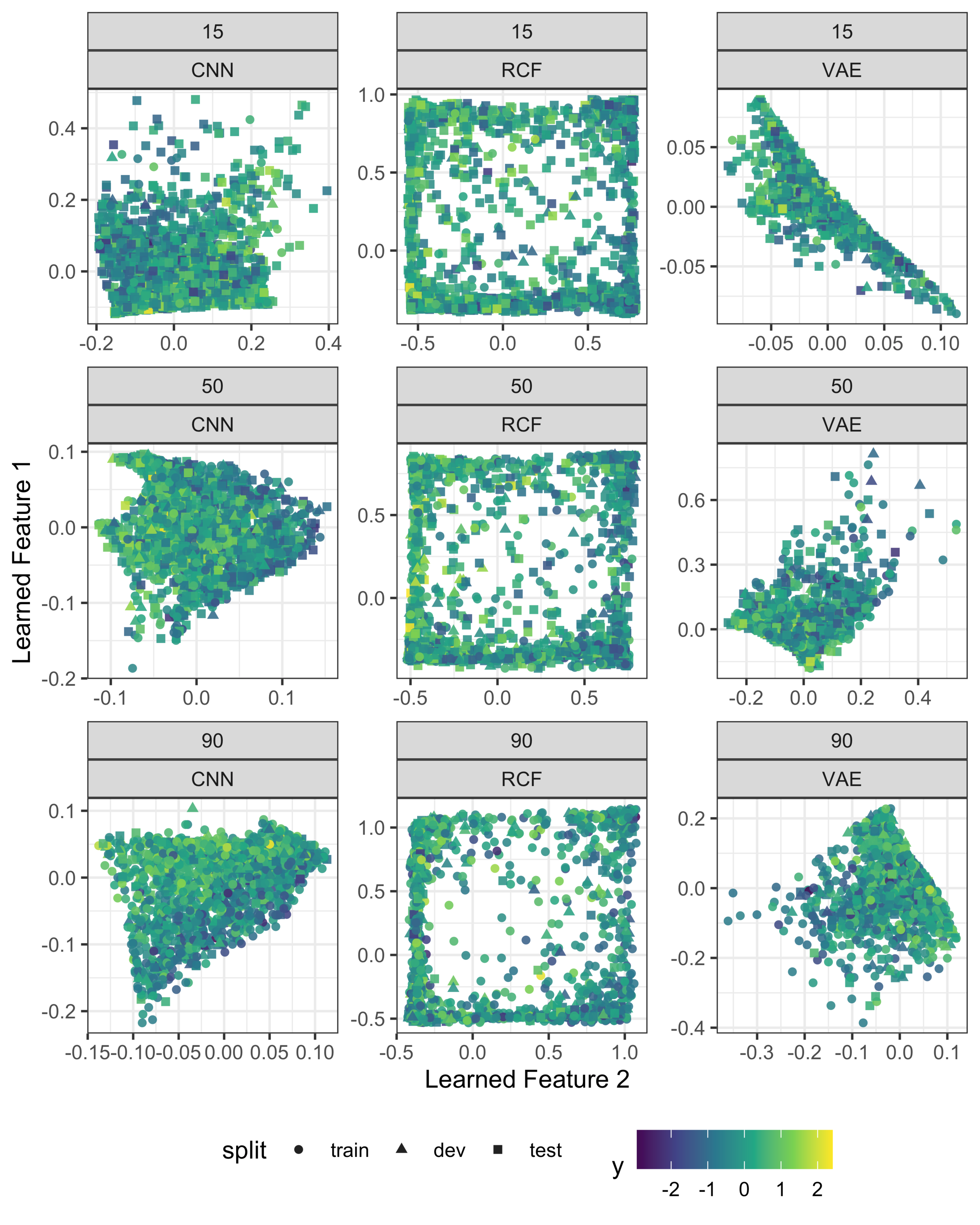}
  \caption{The coordinates of all samples in the first two dimensions after
    aligning the SCA-reduced features. A subset, along with the bootstrap
    glyphs, is shown in Figure \ref{fig:sim_embeddings--sca-1-2}.}
  \label{sim_embeddings-full-sca-1-2}
\end{figure}

\begin{figure}
  \centering
  \includegraphics[width=1.2\textwidth]{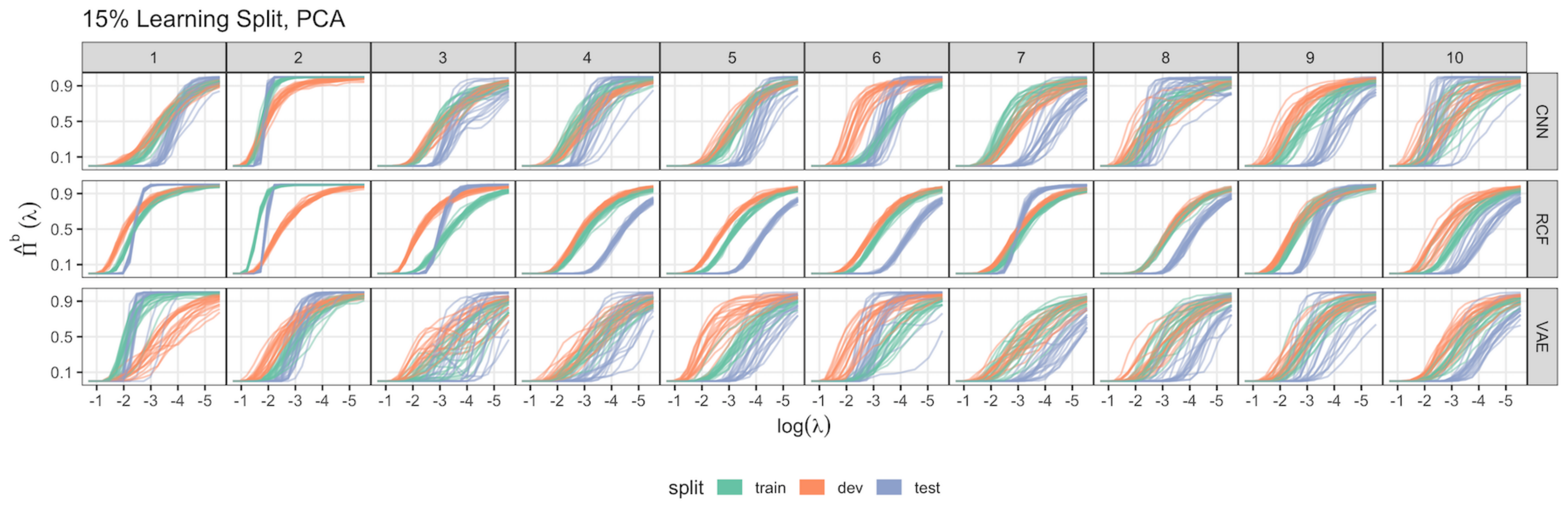}
  \includegraphics[width=1.2\textwidth]{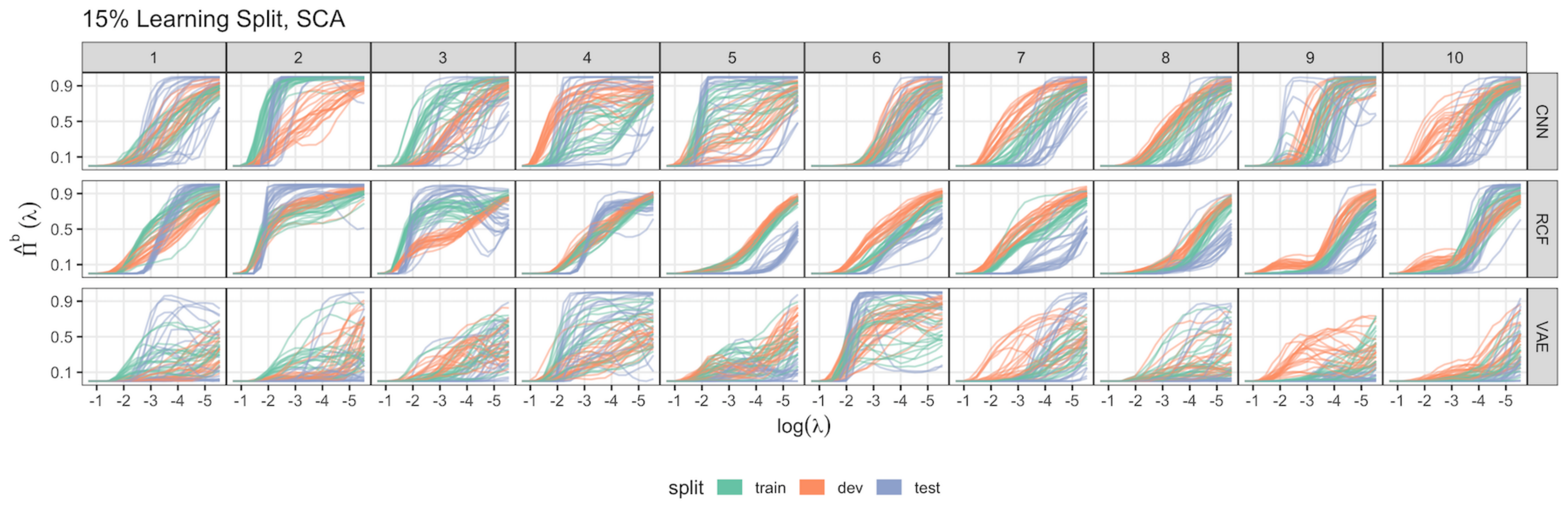}
  \caption{Selection paths for all models when using 15\% of the data for
    feature learning. Most features, especially those from the CNN, have lower
    selection probability than the analogous features at other split sizes.
    Other key characteristics of the selection curves, like the effect of SVD
    vs. SCA reduction, and the stability of RCF features, remain the same.}
  \label{fig:selection_paths15}
\end{figure}

\begin{figure}
  \centering
  \makebox[\textwidth][c]{\includegraphics[width=1.2\textwidth]{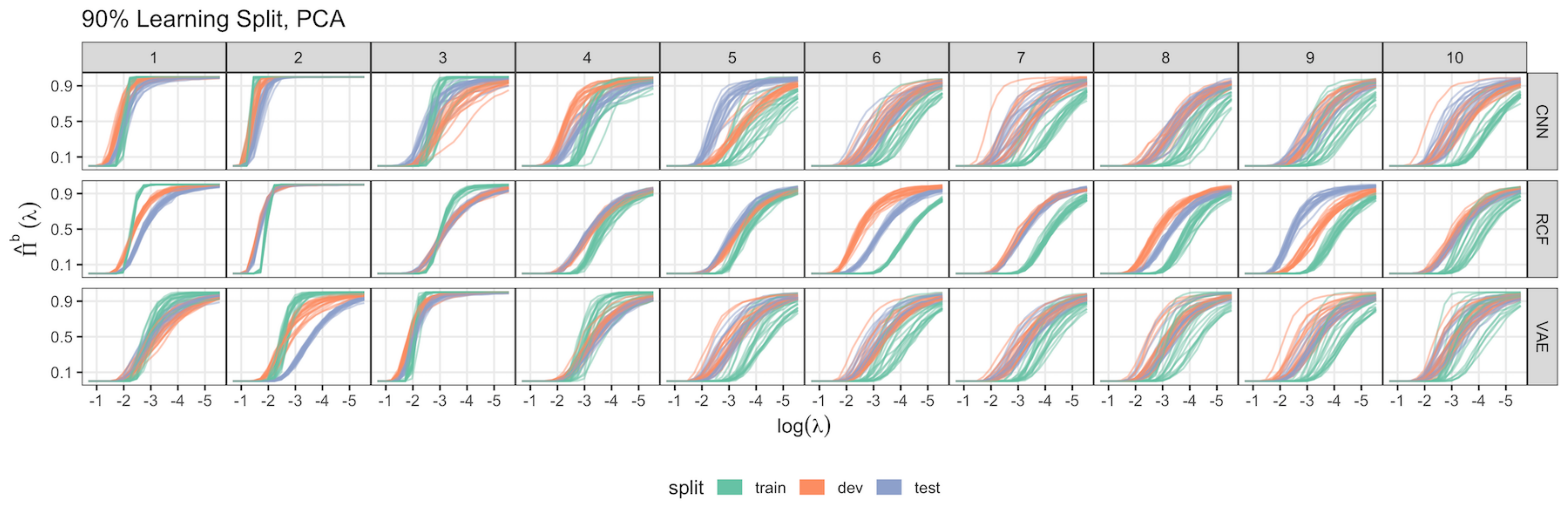}}
  \makebox[\textwidth][c]{\includegraphics[width=1.2\textwidth]{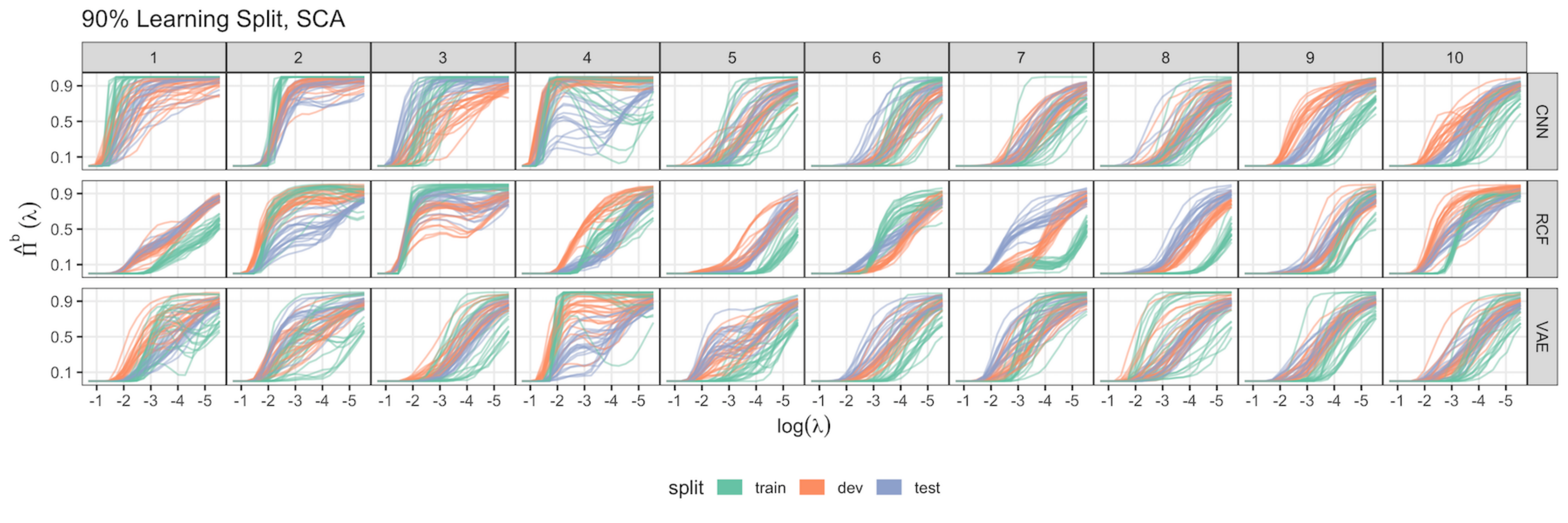}}
  \caption{Selection paths for all models when using 90\% of the data for
    feature learning. The top learned features across several methods are
    strongly related to the response.}
  \label{fig:selection_paths90}
\end{figure}

\begin{figure}
  \centering
  \makebox[\textwidth][c]{\includegraphics[width=1.4\textwidth]{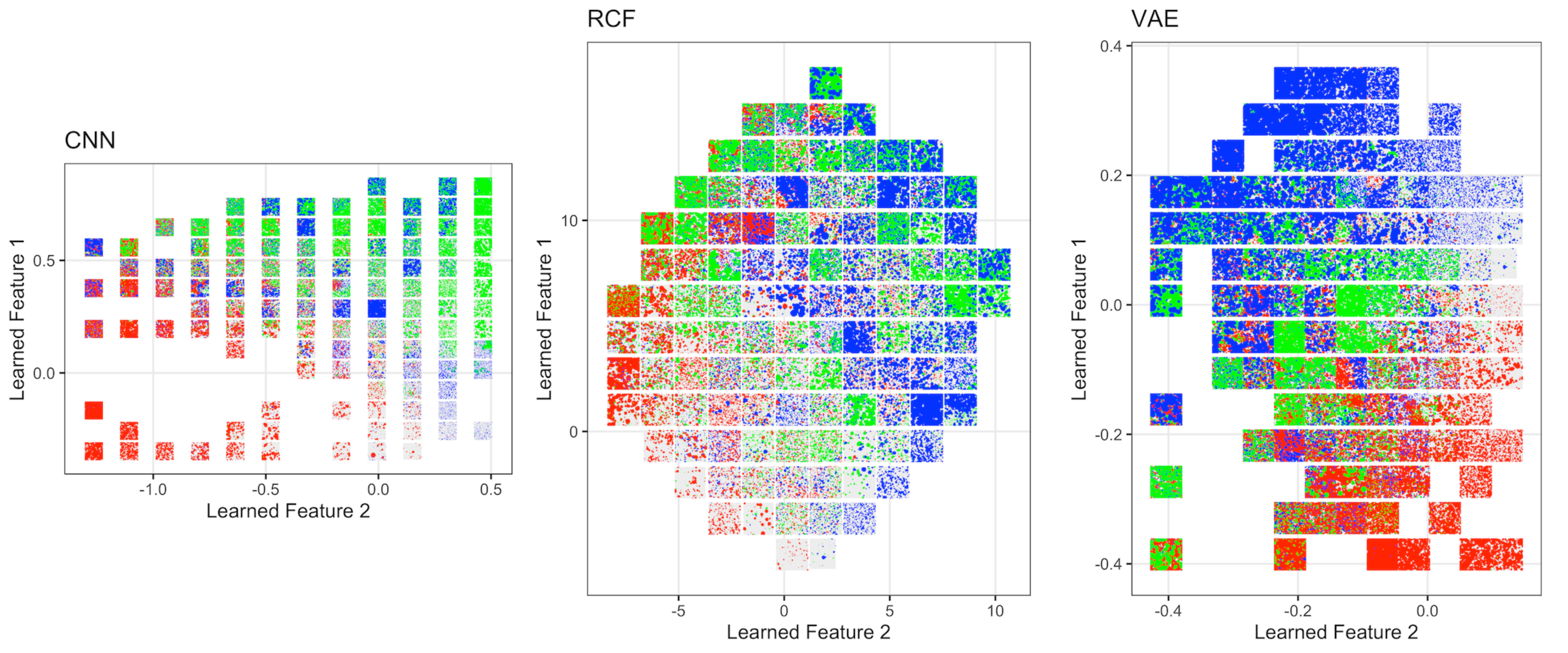}}
  \caption{Patches overlaid on the learned features in Figure
    \ref{fig:sim_embeddings-full-pca-1-2}, to guide interpretation of regions of
    the learned feature space. The figure is read in the same way as Figure
    \ref{fig:tnbc_imagegrid-PCA-1-2}.}
  \label{fig:sim_imagegrid-PCA-15-1-2}
\end{figure}

\begin{figure}
  \centering
  \makebox[\textwidth][c]{\includegraphics[width=1.4\textwidth]{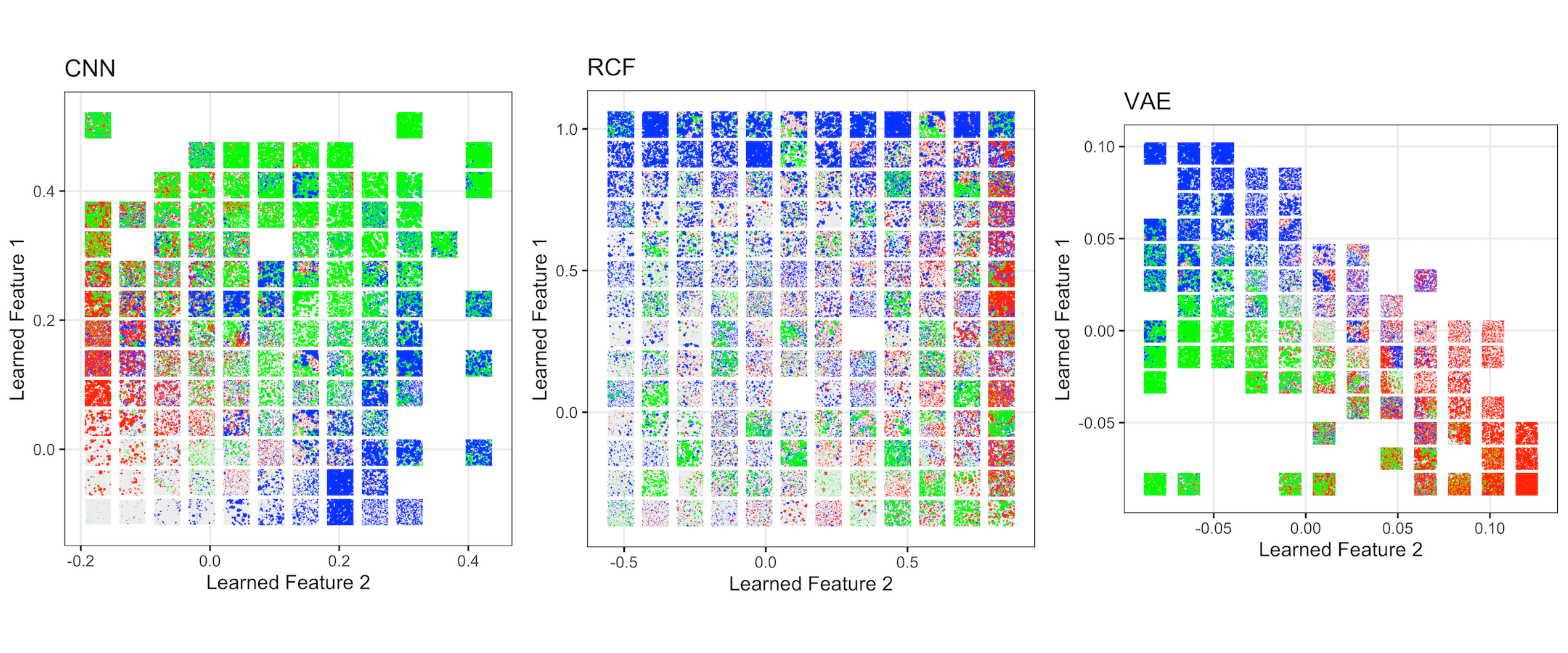}}
  \caption{The analogous figure to \ref{fig:sim_embeddings-full-pca-1-2}, but
    using SCA dimensionality reduction instead.}
  \label{fig:sim_imagegrid-SCA-15-1-2}
\end{figure}

\begin{figure}
\centering
  \includegraphics[width=\textwidth]{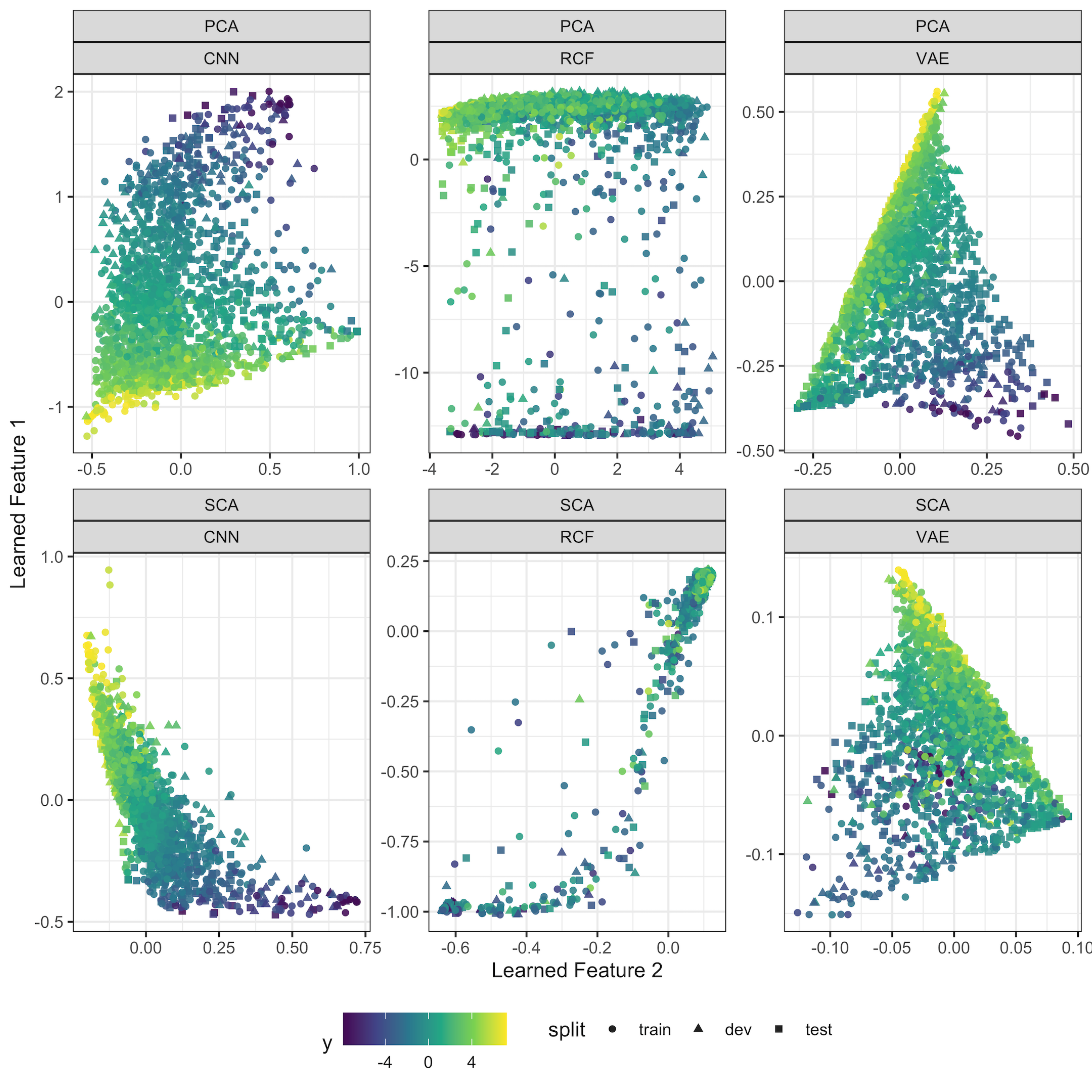}
  \caption{The version of Figure \ref{fig:tnbc_embeddings-1-2} showing all
    samples, and collapsing all glyphs to a point.}
  \label{fig:tnbc_embeddings-full-both-1-2}
\end{figure}

\begin{figure}
  \centering
  \includegraphics[width=\textwidth]{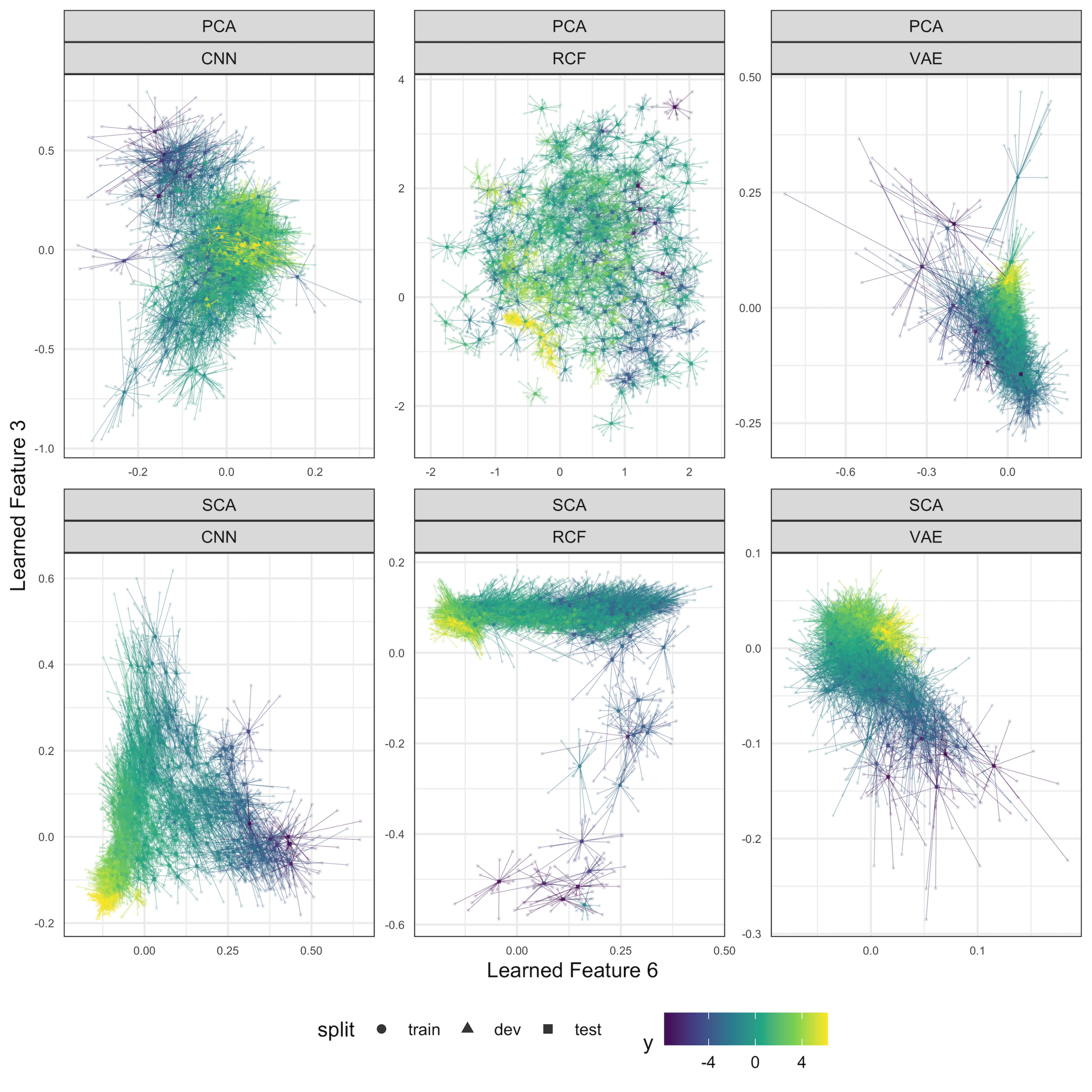}
  \caption{Examples of features that are found to be less important in the TNBC
    dataset application, according to the selection curves in Figure
    \ref{fig:tnbc_selection_paths-2}. Note that the relationship between the
    displayed locations and the measured $\mathbf{y}$ is harder to capture using
    a linear boundary, compared to those in Figure
    \ref{fig:tnbc_embeddings-1-2}.}
  \label{fig:tnbc_embeddings-3-6}
\end{figure}

\begin{figure}
\centering
  \includegraphics[width=\textwidth]{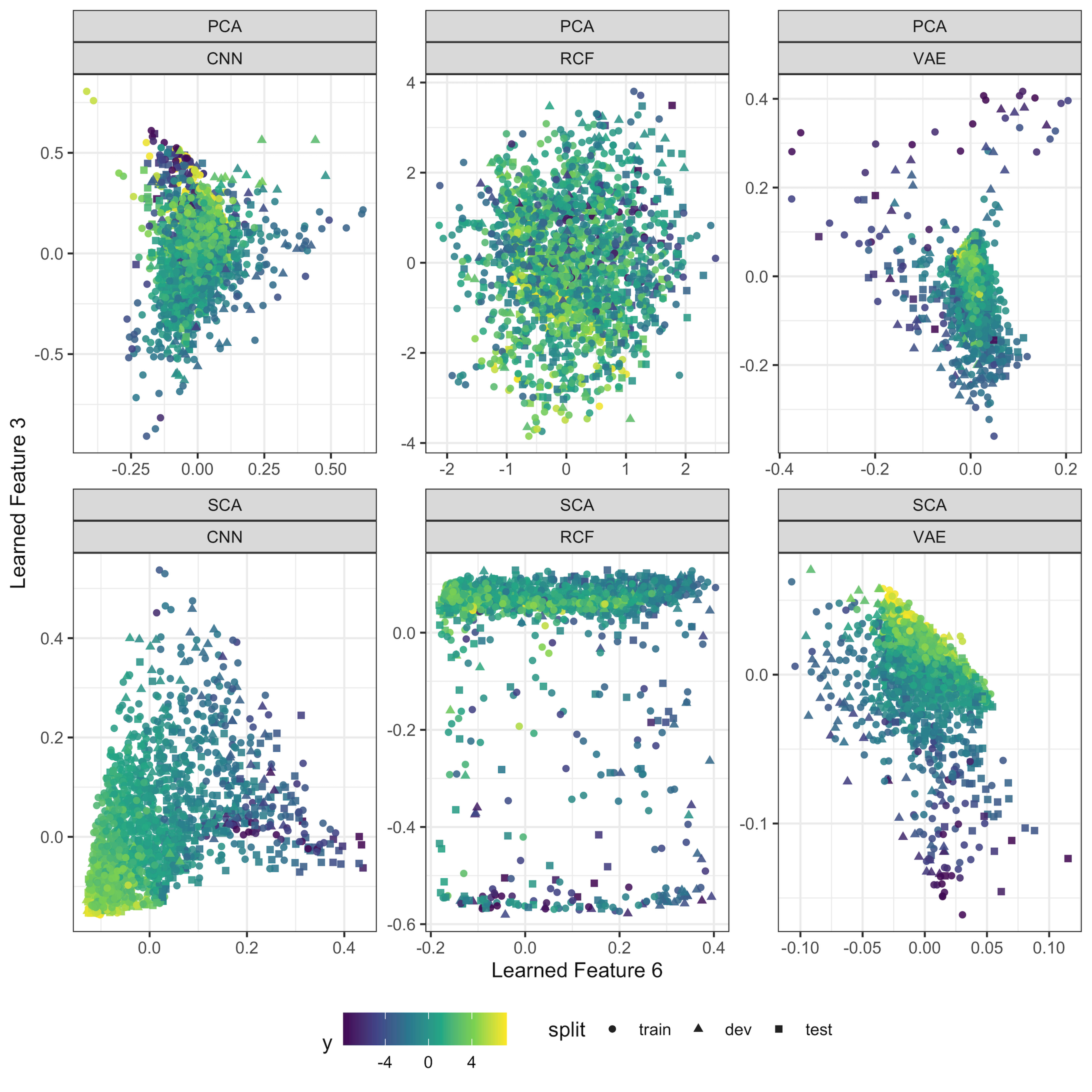}
  \caption{The version of Figure \ref{fig:tnbc_embeddings-3-6} showing all
    samples, and collapsing all glyphs to point.}
  \label{fig:tnbc_embeddings-full-both-3-6}
\end{figure}

\begin{figure}
\centering
  \includegraphics[width=\textwidth]{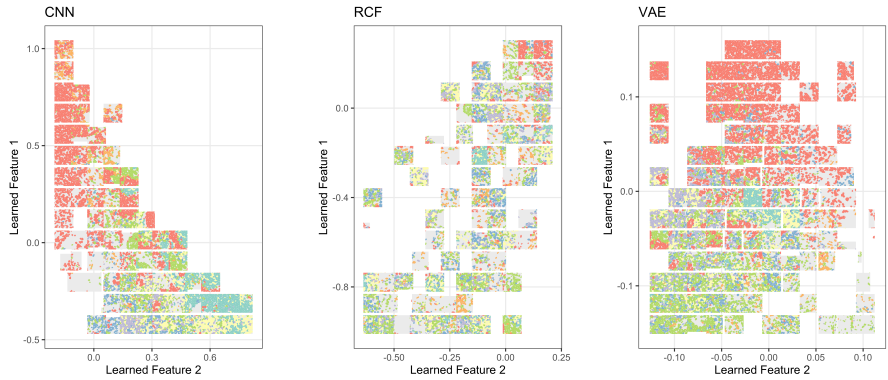}
  \caption{TNBC patches overlaid on the SCA-reduced aligned features, analogous
    to the PCA-reduced view in Figure \ref{fig:tnbc_imagegrid-PCA-1-2}.}
  \label{fig:tnbc_imagegrid-SCA-1-2}
\end{figure}

\begin{figure}
  \centering
  \includegraphics[width=\textwidth]{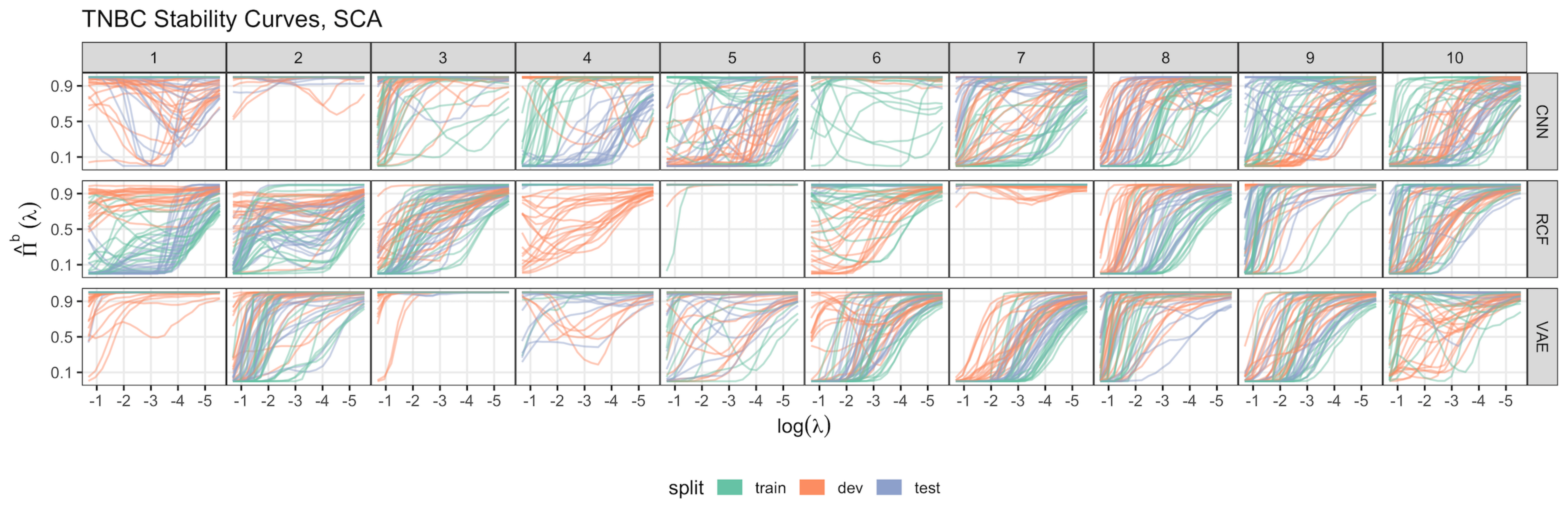}
  \caption{Selection curves for the TNBC dataset application when using SCA for
    dimensionality reduction.}
  \label{fig:tnbc_selection_paths-1}
\end{figure}

\end{document}